\UseRawInputEncoding
\documentclass[12pt]{article}
\usepackage{amsfonts}
\usepackage{mathrsfs}
\usepackage{amsthm}
\usepackage{multirow}
\usepackage{bbding}
\usepackage{amssymb}
\usepackage{amsmath}
\usepackage{graphicx,color}
\usepackage{rotating}
\usepackage{bm}
\usepackage{epstopdf}
\usepackage{cite}
\usepackage{natbib}
\newcommand{\bea}{\begin{eqnarray}}
\newcommand{\eea}{\end{eqnarray}}
\newcommand{\Bea}{\begin{eqnarray*}}
\newcommand{\Eea}{\end{eqnarray*}}
\newcommand{\ba}{\begin{array}}
\newcommand{\ea}{\end{array}}
\newcommand{\bt}{\begin{tabular}}
\newcommand{\et}{\end{tabular}}
\newcommand{\btb}{\begin{table}}
\newcommand{\etb}{\end{table}}
\newcommand{\bc}{\begin{center}}
\newcommand{\ec}{\end{center}}
\newcommand{\beq}{\begin{equation}}
\newcommand{\eeq}{\end{equation}}

\makeatletter

\newcommand{\Rmnum}[1]{\expandafter\@slowromancap\romannumeral #1@}
\makeatother

\begin{document}

\title{\bf 
 Source-Function Weighted-Transfer Learning for Nonparametric Regression with Seemingly Similar Sources}

\author{Lu Lin \thanks{
    The authors gratefully acknowledge \textit{the National Natural Science Foundation of China (grant No. 11971265, 120001318), the National Statistical Science Research Project (grant No. 2022LD03) and the National Key R\&D Program of China (grant No. 2018YFA0703900).}}\hspace{.2cm}\\
    Zhongtai Securities Institute for Financial Studies, Shandong University\\
    and \\
     Weiyu Li \footnote{The corresponding
author. Email: liweiyu@sdu.edu.cn.} \\
    Zhongtai Securities Institute for Financial Studies, Shandong University\\}
  \maketitle

\vspace{-0.7cm}
\begin{abstract} \baselineskip=17pt
The homogeneity, or more generally, the similarity between source domains and a target domain seems to be essential to a positive  transfer learning. In practice, however, the similarity condition is difficult to check and is often violated. In this paper, instead of the popularly used similarity condition, a seeming similarity is introduced, which is defined by a non-orthogonality together with a smoothness. Such a condition is naturally satisfied under common situations and even implies the dissimilarity in some sense. Based on the seeming similarity together with an $L_2$-adjustment, a source-function weighted-transfer learning estimation (sw-TLE) is constructed. By source-function weighting, an adaptive transfer learning is achieved in the sense that it is applied to similar and dissimilar scenarios with a relatively high estimation efficiency. Particularly, under the case with homogenous source and target models, the sw-TLE even can be competitive with the full data estimator.
The hidden relationship between the source-function weighting estimator and the James-Stein estimator is established as well, which reveals the structural reasonability of our methodology. Moreover, the strategy does apply to nonparametric and semiparametric models. The
comprehensive simulation studies and real data analysis can illustrate that the new strategy is significantly better than the competitors.


\end{abstract}
\baselineskip=20pt

\setcounter{equation}{0}
\section{Introduction}

\subsection{Problem setup}
In the processes of generating and collecting data, a common situation is that it is expensive or impossible to obtain adequate training data that can match the feature space and the distribution characteristic of the test data. In such a case, knowledge transfer or transfer learning between task domains would be desirable based on the theoretical belief and/or the available evidence that a transfer learning can improve a learner from one domain by transferring information from some related domains. Up to now, the transfer learning has attracted much attention, and various frameworks and techniques of transfer learning have been proposed in the literature of statistics and machine learning. For the survey papers on transfer learning, see, for example, \citet{storkey2009training}, \citet{pan2009survey}, and \citet{weiss2016survey}; for the applications in the area of computer vision, see \citet{tzeng2017adversarial}, \citet{gong2012geodesic}; for the applications in the area of speech recognition, see \citet{huang2013cross}; for the applications in the area of genre classification, see \citet{choi2017transfer}.

In the familiar methods of transfer learning, a basic condition accompanied by positive transfer learning is
the homogeneity, or more generally, the similarity (or the relatedness) between source domains and the target domain, see, e.g., \citet{yang2015learning}, \citet{li2010negative} and \cite{seah2012combating}.
In practice, however, the homogeneity and similarity conditions are difficult to meet and check, or are often violated. Heterogenous or dissimilar data source is a generally existent phenomenon due to some special causes, such as changes of protocol, advances in technology and periodic variation.
Then, the following issue arises naturally:
\begin{itemize}\item {\it Can the knowledge in heterogenous or dissimilar sources be transferred into the target model to enhance the inferential accuracy of the target model?}\end{itemize}
As far as we know, theoretically, there is no uniform answer to this question.
If the theoretical answer could be ``yes", we would try to develop an adaptive strategy
capable of processing general sources such that the source information can contribute to the inference on the target model whether the source models and the target model are homogenous, or similar or not.

Although the issue above seems to be preposterous, the related topic
could trace back to the nonparametric density  estimation with a parametric start:
\begin{itemize}\item {\it
Can the knowledge in a parametric estimation be transferred into the nonparametric density to enhance the estimation accuracy if the parametric estimation is actually quite crude or away from the true nonparametric density? }\end{itemize}
The answer to the question is in the affirmative for general cases, see, for example, \citet{hjort1995nonparametric} and \citet{naito2004semiparametric}. \citet{lin2009adaptive} extended the method to the case of additive models.

Another early research could date back to  Stein's paradox,  see, e.g., \citet{Stein1956},   \citet{efron1973stein}, \citet{james1992estimation}. The researchers answered ``yes" to the seemingly preposterous question:
\begin{itemize}\item {\it
Can the information about the price of apples in Washington and about the price of oranges in Florida be used to improve an estimate of the price of French wine if the prices are unrelated to each other? }
\end{itemize}
Actually, the shrinkage estimation can achieve this goal via the use of these unrelated data. A famous example of this type estimation is the James-Stein estimator \citep{james1992estimation}, which is shrunk with less estimation risk (see the Supplement Material).

\subsection{Our notion and contribution}

As shown above, in the classical examples of the adaptive density and regression estimators, and James-Stein estimator, the similarity condition between the objects is not always necessary. However, these methods focus only on the traditional scenarios without source domain, which are essentially different from those for transfer learning. Up to now, these classical notions have not been applied to the modern realm of transfer learning. Motivated by these notions, in this paper, we will develop a new strategy and conduct a comprehensive study. Our work is outlined as follows.

{\it a) Seeming similarity condition}.
Instead of the classical similarity condition, a seeming similarity is introduced as a basic condition.
The seeming similarity is defined by a non-orthogonality together with a smoothness. This is a very common condition, and is naturally satisfied under common situations, moreover it even implies dissimilarity in some sense.

{\it b) Source-function weighted-transfer learning}.
Based on such a seeming similarity together with a smoothness, a source-function weighted-transfer learning estimation (sw-TLE) is introduced in this paper. This is a brand-new strategy in the area of transfer learning, and the hidden relationship between the source-function weighting in transfer learning and the James-Stein estimator in linear models is established.
%
This structural reasonability reveals the reason why our method can adaptive to various scenarios and achieve a positive transfer learning.

{\it c) Favorable properties}.
Without the familiar similarity condition, the estimation efficiency of the sw-TLE can be significantly enhanced for the case of seeming similarity and relatively large source. Thus, our method is even applied to the case when the familiar similarity condition is somewhat violated. Particularly, when source models and the target model are quite similar or are equal to each other, the source-function weighted-transfer learning can always enhance the estimation efficiency, and the sw-TLE even can be competitive with the full data estimator.
Moreover, it is surprising that we even can improve the estimation of target function only by artificial data.

Moreover, the strategy does apply to both nonparametric and semiparametric models, and is computationally simple.
However, as a price to pay for the adaptability, there is a certain risk of negative transfer learning.

\subsection{Review of the familiar transfer learning with similarity conditions}

To further motivate our methodological development, we in this subsection recall the familiar similarity conditions and briefly discuss the relevant topics on transfer learning. The similarity conditions used in the existing literature deeply affect the choices of the transfer learning methodologies, the performances and the theoretical conclusions; for theoretical analysis see for example \citet{hanneke2020no}.
Because of the importance of the similarity in the existing methodologies, various rules have been proposed up to now to measure the degree of similarity between source models and the target model. Generally, we can establish the bound of divergence between a source  and the target to evaluate the
level of similarity \citep{mansour2009domain,tian2022transfer}. It is clear that the theoretical properties of the existing methods of transfer learning depend basically on the divergence bounds. However, in the area of parameter-transfer learning, the degree of similarity is evaluated in
the following special ways. The first is to suppose that source models and the target model share some parameters
or prior distributions of the hyperparameters. Then, the transferred knowledge is encoded
into the shared parameters or priors, and by discovering these identical features, the knowledge can be transferred across the tasks. The second way considers the case where the parameters in source models and the target model are close to each other, i.e., the parameters in source models are transferable. The core idea of the second method is to transfer the information from transferable source domains to obtain some rough estimators in the first step, and then use a bias-correction procedure to construct debiased estimators in the second step. The related references on parameter-transfer learning include
\citet{xia2013feature},
 \citet{li2020transfer} and \citet{tian2022transfer}. For feature selection in parametric models see \citet{chen2021transfer}.

In the models with covariates and response variables, it is an often situation where
the marginal distributions of covariates in source models and the target model may be different, but the conditional
distributions of the responses, given the corresponding covariates, are identical.
In this case, the degree of similarity is evaluated by the covariate shift \citep{kpotufe2021marginal}.
Contrarily, the conditional distribution
drift is also a general framework and broadly arises in many practical problems \citep{weiss2016survey,cai2021transfer,reeve2021adaptive}.
In this case, a commonly used method is to formulate
a transfer function $\phi(\cdot)$ together with some propensity constraints. For example, for regression functions $\eta_P(\cdot)$ and $\eta_Q(\cdot)$ respectively in a source model and the target model, the similarity relationship between the functions $\eta_P(\cdot)$ and $\eta_Q(\cdot)$ is described by $\eta_P(\cdot)=\phi(\eta_Q(\cdot))$, see, e.g., \citet{cai2021transfer} and  \citet{reeve2021adaptive}. Such a transfer function $\phi(\cdot)$ can determine the theoretical property, but it is difficult or impossible to formulate its detailed form in practical use.

It can be seen from the above literature review that the similarity is essential in the familiar methodology of transfer learning.
Without these similarity conditions, however, it is difficult to transfer the information across the tasks by the existing methods. Recently, \citet{lin2022correlation} proposed a historical data condition, instead of a similarity condition, to establish a correlation ratio transfer learning. In this method, the historical data condition is employed  to build the relationship between source models and the target model.
Moreover, in the most procedures of the existing transfer learning, data-driven algorithms are often required
to implement the transfer learning, which are computational complex usually. Thus, it is desired to develop adaptive methods together with efficient algorithms to adapt to the case with or without similarity condition.

\subsection{Article structure}

The remainder of the paper is organized as follows. In Section 2, nonparametric regression models with a seeming similarity are defined, an $L_2$-criterion is introduced to adjust an estimator from source start, and then the transfer learning estimator for target regression is suggested, its simple form and extended version are discussed. The asymptotic normalities are established, and the enhancements of convergence rate and estimation efficiency are confirmed theoretically. In Section 3, the method is extended into semiparametric models, the corresponding theories are established as well, and the technique is further extended to the case of multi-source.
Simulation studies and real data analysis are presented in Section 4. Section 5 concludes the paper and discusses possible future work. The Supplement Material includes the following materials: the smoothness of the adjustment function, some regularity conditions, some additional theoretical conclusions, the proofs of the theorems and corollaries, and the extended simulation studies and the relationship between our estimator and the James-Stein estimator.

\setcounter{equation}{0}
\section{Methodology for seemingly similar nonparametric models}

In this section, we consider a nonparametric target regression combined with a single nonparametric source. The cases of multi-source and semiparametric regressions will be investigated in the next section.

\subsection{Models}

The models under study in this section are the following nonparametric regressions:
\begin{equation}\begin{split}\label{model-P}&\mbox{(P-model)} \ \ \ Y_i^P=r^P(X_i^P)+
\varepsilon_i^P,i=1,\cdots,n^P,\\ & \mbox{(Q-model)} \ \ \ Y_i^Q=r^Q(X_i^Q)+\varepsilon_i^Q,i=1,
\cdots,n^Q,\end{split}\end{equation}
where $r^P(x)$ and $r^Q(x)$ are unknown nonparametric regression functions. In this paper, it is supposed that $x$ is a scale variable for simplicity. Under multivariate case, the method is similar.
When the above models are fixed designed, we suppose without loss of generality that regression functions $r^P(x)$ and $r^Q(x)$ are defined on the common interval $[0,1]$, and the design variables in data sets $D^P=\{(X_i^P,Y_i^P),i=1,\cdots,n^P\}$ and $D^Q=\{(X_i^Q,Y_i^Q),i=1,\cdots,n^Q\}$
satisfy $0\leq X_1^P\leq \cdots \leq X^P_{n^P}\leq 1$ and $0\leq X_1^Q\leq \cdots \leq X^Q_{n^Q}\leq 1$. In this case, the model errors have the expectations and variances as $\mathbb{E}[\varepsilon_i^P]=0$, $\mathbb{V}[\varepsilon_i^P]=\sigma_P^2$, $\mathbb{E}[\varepsilon_i^Q]=0$ and $\mathbb{V}[\varepsilon_i^Q]=\sigma_Q^2$.
Under random design models, $D^P$ and $D^Q$ are random samples. In this case, we suppose without loss of generality that regression functions $r^P(x)$ and $r^Q(x)$ are defined on a common interval $[L,U]$.
In random design models, the model errors satisfy
$\mathbb{E}[\varepsilon_i^P|X_i^P]=0$, $\mathbb{V}[\varepsilon_i^P|X_i^P]=\sigma_P^2$, $\mathbb{E}[\varepsilon_i^Q|X_i^Q]=0$ and $\mathbb{V}[\varepsilon_i^Q|X_i^Q]=\sigma_Q^2$.
In this paper, we denote ``$\mbox{P}=\mbox{Q}$" if $r^P(x)=r^Q(x)=r(x)$ for all $x$, and $\sigma^2_P=\sigma^2_Q=\sigma^2$.

P-model is regarded as a source model, from which most training data are collected, and Q-model is thought of as a target model, about which  we wish to make inference.
In this paper, our purpose is to estimate the target function $r^Q(x)$. Note that the two regression functions $r^P(x)$ and $r^Q(x)$ may be distinct from each other, and the difference between them may be relatively large. We in this paper use the following non-orthogonality condition \begin{equation}\begin{split}\label{(condition)} \langle r^P,r^Q \rangle_{h^Q}(x)\neq 0 \mbox{ as }h^Q\rightarrow 0 \mbox{ for any } x\end{split}\end{equation} together with a smoothness condition (given in condition C3) to describe a seeming similarity between $r^P(x)$ and $r^Q(x)$. Here $\langle r^P,r^Q \rangle_{h^Q}(x)$ is the (local) inner product between $r^P(x)$ and $r^Q(x)$ at a target point $x$, defined by
$\langle r^P,r^Q \rangle_{h^Q}(x)=\int_0^1 K_{h^Q}(t-x)r^Q(t){r}^P(t)dt$ under fixed design model, or defined by $\langle r^P,r^Q \rangle_{h^Q}(x)=\mathbb {E}[K_{h^Q}(X^Q-x)r^Q(X^Q){r}^P(X^Q)]$ under random design model, where $K_{h^Q}(\cdot)=K(\cdot/h^Q)/h^Q$,  $K(\cdot)$ is a kernel function and $h^Q$ is the bandwidth depending on the sample size of Q-model.
Actually, the condition (\ref{(condition)}) is naturally satisfied under common situation, because it only excludes the case of the absolute dissimilarity --- the orthogonality between $r^P(x)$ and $r^Q(x)$. The above seeming similarity condition is totally different the common similarity condition (see, e.g., \citealp{cai2022transfer}), and is even implies the dissimilarity in some sense.
Thus, the classical similarity conditions may be violated. In this case, it is difficult to achieve a positive transfer learning by the existing methods.

\subsection{The method for fixed design model}

We first focus on the fixed design model. According to the fixed design points,
we set $s_0^P=0,s_i^P=(X_i^P+X_{i+1}^P)/2$ for $i=1,\cdots,n^P-1$, and $s_{n^P}=1$, and, similarly, set $s_0^Q=0,s_i^Q=(X_i^Q+X_{i+1}^Q)/2$ for $i=1,\cdots,n^Q-1$, and $s_{n^Q}=1$. To employ the information of P-model in the procedure of estimating the target function $r^Q(x)$, we first construct a nonparametric estimator of $r^P(x)$ based on P-model. As an example, the Gasser-M\"uller (G-M) estimator is chosen  specifically for this purpose, which is defined by
$\widehat {r}^P(x)=\sum_{i=1}^{n^P}Y_i^P\int_{s_{i-1}^P}^{s_i^P}K_{h^P}(t-x)dt
$.
Although $\widehat {r}^P(x)$ is a consistent estimator of $r^P(x)$, it may be far away from the target function $r^Q(x)$ because $r^P(x)$ and $r^Q(x)$ may be greatly different. 
We then regard $\widehat {r}^P(x)$ as an initial estimator of $r^Q(x)$ and try to adjust it to the following form:
\begin{equation}\label{(general)}f(\widehat {r}^P(x),x)\mbox{ with } f(0,x)=0,\end{equation} where the function $f(\cdot,\cdot)$ is required to be specified. Here we need the condition $f(0,x)=0$ to delete the intercept when the function is expressed by basis functions because the intercept dose not contain any information of $\widehat r^P(x)$; see (\ref{(polynomial)}) and the Supplement Material.

\subsubsection{Linear framework}

As a major choice,
the function $f(\cdot,\cdot)$ is set as the following linear framework:
\begin{equation}\label{(linear)}f(\widehat {r}^P(x),x)=\widehat {r}^P(x)\xi(x),\end{equation}
where $\xi(x)$ could be regarded as an adjustment factor to be estimated.
In this subsection, we focus on the above linear framework. The general case of $f(\cdot,\cdot)$ will be investigated in the next subsection.
Here, we introduce the following strategy: the function $\xi(x)$ in (\ref{(linear)}) is determined by minimizing
the local $L_2$-criterion:
$\int_0^1 K_{h^Q}(t-x)\{r^Q(t)-\widehat {r}^P(t)\xi(x)\}^2dt$ at a fixed target point $x$.
We then get the solution of $\xi(x)$ as
\begin{equation}\begin{split}\label{(solution)}\xi_{h^Q}(x)=\frac{\int_0^1 K_{h^Q}(t-x)r^Q(t)\widehat {r}^P(t)dt}{\int_0^1 K_{h^Q}(t-x)
(\widehat {r}^P(t))^2dt}.\end{split}\end{equation} The adjustment factor $\xi_{h^Q}(x)$ in (\ref{(solution)}) could be thought of as the normalized inner product between $\widehat r^P(x)$ and $r^Q(x)$ according to the definition of inner product in (\ref{(condition)}).
In the above, however, the numerator of the adjustment factor $\xi_{h^Q}(x)$ depends on unknown function $r^Q(t)$. We propose the following method to approximate it for the case where the fixed design points $X_1^Q, \cdots, X^Q_{n^Q}$ satisfy the quasi-uniform condition. Here, the quasi-uniformity means that the difference among $|X^Q_i-X^Q_{i-1}|,i=2,\cdots,n^Q$, is small; the detailed definition of the quasi-uniformity will be given in condition C2.
With the quasi-uniform condition, if the function $r^Q(x)$ is smooth, the numerator of $\xi_{h^Q}(x)$ can be approximated by
\begin{equation}\label{integral}\begin{split}
&\int_0^1 K_{h^Q}(t-x)r^Q(t)\widehat {r}^P(t)dt
\approx \sum_{i=1}^{n^Q}Y_i^Q\int_{s_{i-1}^Q}^{s_i^Q}K_{h^Q}(t-x)\widehat {r}^P(t)dt.\end{split}\end{equation}
We thus get the estimated adjustment factor as
\begin{equation}\begin{split}\label{(projection)}\widehat\xi(x)=
\frac{\sum_{i=1}^{n^Q}Y_i^Q\int_{s_{i-1}^Q}^{s_i^Q}K_{h^Q}(t-x)\widehat {r}^P(t)dt}{\int_0^1 K_{h^Q}(t-x)
(\widehat {r}^P(t))^2dt}.\end{split}\end{equation}
Consequently, based on an initial estimator $\widehat {r}^P(x)$ from the P-model, the transfer learning estimator of $r^Q(x)$ for the Q-model can be chosen as
\begin{equation}\label{estimate-A}\widehat {r}_a^Q(x)=\widehat {r}^P(x)\widehat\xi(x).\end{equation} Approximately, the estimator $\widehat {r}_a^Q(x)$ is a (locally) weighted sum of $Y_i^Q$ with the source-function $\widehat{r}^P(X_i^Q)$ as the weights (see (\ref{(projection)})). We then call the estimator $\widehat {r}_a^Q(x)$ as {\it the source-function weighted-transfer learning estimator}, denoted by sw-TLE for short. The above framework of source-function weighting could be regarded as an analog of the James-Stein estimator; the details for verifying this point of view will be given in the Supplement Material.
In the estimator, the empirical choices of the bandwidths can be determined by cross validation (CV) criterion. Let $\widehat {r}^Q_{a(-i)}(x)$ be the leave-one-out form of $\widehat {r}_a^Q(x)$ from the data set $D^Q$. Then the choices of the bandwidths $h^P$ and $h^Q$ can be implemented by minimizing the following CV criterion:
\begin{equation}\label{(CV)}\mathbb{CV}(h^P,h^Q)=\frac{1}{n^Q}
\sum_{j=1}^{n^Q}\left(Y_i^Q-\widehat {r}^Q_{a(-i)}(X_i^Q)\right)^2.\end{equation}


Before investigating the theoretical properties of
the estimator (\ref{estimate-A}), we first look at its structural reasonability:
\begin{itemize}\item[(i)] {\it Adaptability and smoothness.} If the source regression function $r^P(x)$ captures the main features of the shape of the target regression function $r^Q(x)$, for example $r^P(x)\approx cr^Q(x)$ for all $x\in[0,1]$ with a constant $c$, then the adjustment factor $\xi_{h^Q}(x)$ defined in (\ref{(solution)}) satisfies  $\xi_{h^Q}(x)\approx 1/c$ for all $x\in[0,1]$, a smooth constant function, approximately. Furthermore, in the Supplement Material we will verify that, in many cases, $\xi_{h^Q}(x)$ is smoother than $r^Q(x)$. Generally, for general adjustment function $f(\cdot,\cdot)$ given in (\ref{(general)}), it will be verified in the Supplement Matrical that its ideal choice is an identity function, approximately, implying that it is smoother than $r^Q(x)$. These show that the adjustment functions $\xi_{h^Q}(\cdot)$ and $f(\cdot,\cdot)$ are easier to estimate nonparametrically, with an accelerated convergence rate.
\item[(ii)] {\it The framework of the weighted sum of all the data.}
It will be shown in the proof of Theorem 2.1 that actually the sw-TLE can be expressed approximately as a weighted sum of all the data $Y^P_i$ and $Y^Q_i$:
\begin{equation}\begin{split}\label{(decomposition)}\widehat {r}_a^Q(x)\approx r^Q(x)+
\frac{r^Q(x)}{r^P(x)}\left(\widehat r^P(x)-r^P(x)\right) +r^P(x)\left(\sum_{i=1}^{n^Q}
w_i(x)Y^Q_i
-\xi^*_{h^Q}(x)\right),\end{split}\end{equation}
where $w_i(x)=
\frac{\int_{s_{i-1}^Q}^{s_i^Q}K_{h^Q}(t-x)  {r}^P(t)dt}{\int_0^1 K_{h^Q}(t-x)
( {r}^P(t))^2dt}$ and $\xi^*_{h^Q}(x)
=\frac{\int_0^1 K_{h^Q}(t-x)r^Q(t){r}^P(t)dt}{\int_0^1 K_{h^Q}(t-x)
({r}^P(t))^2dt}$.
Then the asymptotic normality can be achieved.
\end{itemize}
The above structural reasonability reveals the hidden reason why our method can achieve a positive transfer learning.

Now we establish the detailed asymptotic theory. The regularity conditions for the kernel function and bandwidths are listed in the Supplement Material. Particularly, the following special condition is needed:

\begin{itemize}
\item[C1.] For all $x$, the regression functions
$r^P(x)$ and $r^Q(x)$
have the second-order continuous and bounded derivatives, $r^P(x)\neq 0$, and the inner product satisfies $\langle r^P,r^Q \rangle_{h^Q}(x)\neq 0$ as $h^Q\rightarrow 0$.
\end{itemize}
In C1, the condition of $r^P(x)\neq 0$ seems to be unreasonable, but it can be removed from C1. For example, if $r^P(x)= 0$ for a point $x$, we transform $Y^P$ into $\widetilde Y^P=Y^P+a$ for a constant $a\neq 0$. By this treatment, P-model can be recast as $\widetilde Y^P=\widetilde r^P(x)+\varepsilon$ with $\widetilde r^P(x)=r^P(x)+a\neq 0$ for all $x$ if $a$ is large enough.
The most important condition  in C1 is the (local) non-orthogonality $\langle r^P,r^Q \rangle_{h^Q}(x)\neq 0$. We need the non-orthogonality to ensure the transferability of P-model because the transfer learning method aforementioned is based on a projection. This condition indicates that $r^P(x)$ and $r^Q(x)$ is not absolutely dissimilar.

\noindent{\bf Theorem 2.1.} {\it For the fixed design model, in addition to the condition C1, and the regularity conditions S1 and S2 in the Supplement Material, suppose that the following quasi-uniform condition holds:
\begin{itemize}
\item[C2.] The data points $D^P$ and $D^Q$ are fixedly designed respectively by two design functions $\varphi^P(x)>0$ and $\varphi^Q(x)>0$, satisfying $s_i^P -s_{i-1}^P=1/(\varphi^P(X^P_i)n^P)+o(1/n^P)$ and $s_i^Q -s_{i-1}^Q=1/(\varphi^Q(X^Q_i)n^Q)+o(1/n^Q)$ for all $i$, where the given design functions have the second-order continuous and bounded derivatives for all $x\in[0,1]$.
   \end{itemize} Furthermore, suppose that the sample sizes and bandwidths satisfy $h^P=o((n^P)^{-1/5})$ and $h^Q=o((n^Q)^{-1/5})$, and $n^Ph^P/(n^Qh^Q)\rightarrow \tau$ for some constant $0\leq\tau\leq\infty$. Then the sw-TLE (\ref{estimate-A}) has the following asymptotic normality:

(i)  for $ 0<\tau<\infty$ and $ x\in(0,1)$,
\begin{equation*}\begin{split}&\sqrt{n^Ph^P+n^Qh^Q}
\left(\widehat {r}_a^Q(x)-r^Q(x)\right)\\&\stackrel{d}\rightarrow
N\left(0,\left(1+\frac{1}{\tau}\right)\left(\frac{2r^Q(x)
}{r^P(x)}\right)^2
\frac{\sigma^2_P}
{\varphi^P(x)}\int_{0}^{1}K^2(t)dt+\frac{\sigma^2_Q(1+\tau)}
{\varphi^Q(x)}\int_{0}^{1}K^2(t)dt
\right);\end{split}\end{equation*}

(ii) for $\tau=\infty$ and $ x\in(0,1)$,
\begin{equation*}\begin{split}\sqrt{n^Qh^Q}
\left(\widehat {r}_a^Q(x)-r^Q(x)\right)\stackrel{d}\rightarrow
N\left(0,\frac{\sigma^2_Q}
{\varphi^Q(x)}\int_{0}^{1}K^2(t)dt
\right)\end{split};\end{equation*}

(iii)  for $ \tau=0$ and $ x\in(0,1)$,
\begin{equation*}\begin{split}\sqrt{n^Ph^P}
\left(\widehat {r}_a^Q(x)-r^Q(x)\right)\stackrel{d}\rightarrow
N\left(0,\left(\frac{2r^Q(x)
}{r^P(x)}\right)^2
\frac{\sigma^2_P}
{\varphi^P(x)}\int_{0}^{1}K^2(t)dt
\right).\end{split}\end{equation*}

}

The proof of the theorem depends mainly on the decomposition in (\ref{(decomposition)}), a weighted sum of all $Y_i^P$ and $Y_i^Q$ (see the Supplement Material).
The quasi-uniform condition C2 in the theorem can be easily implemented. For example, the simplest
case for model Q is that the design function is chosen as $\varphi^Q(x)\equiv 1$, resulting in $s_i^Q -s_{i-1}^Q=1/n^Q$ for all $i$. Generally, for a given design function $\varphi^Q(x)>0$, we set $F(x)=\int_0^x\varphi^Q(t)dt$ and $P(x)=F^{-1}(x)$, then the design points $X_i^Q=P\left(\frac{i-1/2}{n^Q}\right)+o(1)$ for all $i$ satisfy the quasi-uniform condition. It is worth pointing out that under our models, in addition to the possible dissimilarity between regression functions $r^P(x)$ and $r^Q(x)$, the design functions $\varphi^P(x)$ and $\varphi^Q(x)$ may be dissimilar.
In addition, in the asymptotic normality above, the condition $x\in(0,1)$ is not a necessary constraint; that is, we use it only for simplicity
of presentation, due to the boundary effect of the G-M kernel estimators used in the previous subsection.

In the theorem, we use the under-smoothing condition $h^P=o((n^P)^{-1/5})$
and $h^Q=o((n^Q)^{-1/5})$ to reduce the asymptotic bias and then to get a
concise expression of the asymptotic normality. In general nonparametric kernel estimators, the bandwidths are set to satisfy the optimal bandwidth condition $h^P=O((n^P)^{-1/5})$
and $h^Q=O((n^Q)^{-1/5})$. With the optimal bandwidths, the sw-TLE has the asymptotic bias of order $O_p((h^P)^2+(h^Q)^2)$. Because the representation of the asymptotic bias is complicated in our method (see the proof of Theorem 2.1), the details on the asymptotic bias are omitted here.

For better understanding the theorem, we give the following remark.

\noindent{\bf Remark 2.1.} {\it
\begin{itemize}
\item[(i)]
Under the case of $0<\tau<\infty$, it is possible to enhance the estimation efficiency. For example, by the condition, we have
    $$\mathbb{V}_{\widehat {r}^Q_a}\approx \left(\frac{\sigma^2_P\int_0^1\left(\frac{2r^Q(x)
}{r^P(x)}\right)^2
\frac{1}
{\varphi^P(x)}dx}{\tau}+\sigma^2_Q\int_0^1\frac{1}
{\varphi^Q(x)}dx\right)\frac{\int_{0}^{1}K^2(t)dt}{n^Qh^Q},$$
where $\mathbb{V}_{\widehat {r}^Q_a}=\int_0^1\mathbb{V}[\widehat {r}_a^Q(x)]dx$, the mean integrated variance of $\widehat r_a^Q(x)$. Thus, we can enhance the estimation efficiency by moderately increasing $\tau$, i.e., by moderately increasing the sample size of P-model. However, it is impossible to excessively enhance the estimation efficiency by
immoderately increasing data in P-model (see the second result of the theorem).

\end{itemize}
}

For convergence rate, we have the following further explanation.

\noindent{\bf Remark 2.2.}

{\it
\begin{itemize}
\item[]
For establishing the asymptotic normality, we suppose $h^Q\rightarrow 0$. Sometimes this condition is not necessary. It is known that $\xi_{h^Q}(x)$ will be a constant function approximately when $\theta_o^P(x)\approx c\theta_o^Q(x)$ for a constant $c\neq0$. In this case, the bandwidth $h^Q$ should be a large constant, resulting in a convergence rate of order $O_p(\sqrt{n^Ph^P+n^Q})$, a ``semiparametric rate" combining nonparametric rate in model P with parametric rate in model Q.

\end{itemize}}

Next, we discuss the relative efficiency.
We first compare the sw-TLE $\widehat {r}_a^Q(x)$ with the local data G-M estimator $\widehat r^Q(x)$ obtained only by the data from model Q.
As shown by Theorem 2.1, under the condition of under-smoothing, the asymptotic biases can be ignored relative to the asymptotic variance. The relative efficiency of the estimator $\widehat {r}_a^Q(x)$ is then defined by $\mathbb{RE}[\widehat {r}_a^Q,\widehat r^Q(x)]=\frac{\mathbb{V}^Q}{\mathbb{V}_{\widehat {r}^Q_a}},$ where $\mathbb{V}_{\widehat {r}^Q_a}=\int_0^1\mathbb{V}[\widehat {r}_a^Q(x)]dx$ and $\mathbb{V}^Q=\int_0^1\mathbb{V}[\widehat r^Q(x)]dx$. Denote by $\widetilde h^Q$ the bandwidth used in the local data G-M estimator $\widehat r^Q(x)$.

\noindent{\bf Corollary 2.2.} {\it In addition to the conditions of Theorem 2.1, suppose $h^Q/\widetilde h^Q\rightarrow \rho$. Then, for the case of $ 0<\tau<\infty$, the relative efficiency can be expressed asymptotically as
$$
\mathbb{RE}[\widehat {r}_a^Q(x),\widehat r^Q(x)]=\frac{\rho\tau\sigma_Q^2\int_0^1
\frac{1}{\varphi^Q(x)}dx}{\sigma_P^2\int_0^1\left(\frac{2r^Q(x)}{r^P(x)}
\right)^2\frac{1}{\varphi^P(x)}dx+\tau\sigma_Q^2\int_0^1
\frac{1}{\varphi^Q(x)}dx}.$$
Consequently, we have $\mathbb{RE}[\widehat {r}_a^Q(x),\widehat r^Q(x)]> 1$ if and only if the following condition holds:
\begin{itemize}
\item[C3.] $\rho>1$ and $\tau>\frac{\sigma_P^2\int_0^1\left(\frac{2r^Q(x)}{r^P(x)}
\right)^2\frac{1}{\varphi^P(x)}dx}{\sigma_Q^2\left (\rho-1\right)\int_0^1
\frac{1}{\varphi^Q(x)}dx}.$
\end{itemize}
}

For the corollary, we have the following explanation.

\noindent{\bf Remark 2.3.} {\it Actually, the condition ``$\rho>1$" in C3 could be thought of as a smoothness condition of the adjustment function $\xi_{h^Q}(x)$.
As stated before, $\xi_{h^Q}(x)$ is smoother than $r^Q(x)$ in many cases (for more details see the Supplement Material). Thus the bandwidth $h^Q$ should be larger than $\widetilde h^Q$ usually, implying that the condition ``$\rho>1$" is commonly satisfied. For example, under some regularity conditions, the bandwidths can be chosen as $h^Q=\frac{c}{\left(\int_a^b\left(\ddot \xi_o(x)\right)^2dx
\right)^{\alpha_1}}\frac{1}{(n^Q)^{\alpha_2}}$ with $\xi_o(x)=\frac{r^Q(x)}{r^P(x)}$,
and $\widetilde h^Q=\frac{c}{\left(\int_a^b\left(\ddot{r}^Q(x)\right)^2dx
\right)^{\alpha_1}}\frac{1}{(n^Q)^{\alpha_2}}$ for some constants $c>0$ and $0<\alpha_1,\alpha_2< 1$; see, e.g.,
\citet{hart2013nonparametric}. With the choice, according to the argument in the Supplement Material, we have $\rho=\frac{\left(\int_a^b\left(\ddot{r}^Q(x)\right)^2dx
\right)^{\alpha_1}}{\left(\int_a^b\left(\ddot \xi_o(x)\right)^2dx
\right)^{\alpha_1}}>1$ in many cases. Particularly, when $r^P(x)$ is strictly similar to $r^Q(x)$, $h^Q$ is a positive constant, leading to
$\rho\rightarrow \infty$.
Therefore the result of the corollary ensures that our transfer learning can enhance the estimation efficiency if the value of $\tau$ is relatively large (i.e., the sample size of P-model is moderately large). However, if the value of $\tau$ is relatively small, the relative efficiency may be smaller than 1, consequently, the method could result in a negative transfer learning. }

Moreover, we have the following particular conclusion on artificial data.

\noindent{\bf Corollary 2.3.} {\it Under the conditions of Theorem 2.1, for the case of $0<\tau<\infty$, if $\sigma_P=0$ and $\rho>1$, it always holds that
$\mathbb{RE}[\widehat {r}_a^Q(x),\widehat r^Q(x)]> 1.$}

The corollary presents an extreme situation. It shows that when the auxiliary data come from an extremely accurate regression (i.e., from a strict function relationship): $Y_i^P=r^P(X_i^P)$, the sw-TLE can always enhance the estimation efficiency if $\rho>1$. This implies a seemingly
counter-intuitive conclusion that we can use artificial data to improve the estimation of the target function. However, the artificial data should be well-chosen to satisfy the condition $\rho>1$, implying that it is difficult to realize.

Finally, we compare the sw-TLE $\widehat {r}_a^Q(x)$ with the full data G-M estimator $\widehat r(x)$ obtained by the full dataset $D^P\cup D^Q$. In this case, the full sample size is $n=n^P+n^Q$, and the full design point set is denoted by $D_X=\{X_i, i=1,\cdots,n\}=\{X_i^P,i=1,\cdots,n^P\}\cup\{X_i^Q,i=1,\cdots,n^Q\}$, satisfying $X_1\leq\cdots\leq X_n$. Furthermore, we use $h$ to denote the bandwidth in the full data estimator $\widehat r(x)$, and use $\varphi(x)$ to denote the design function for full design data set $D_X$.
It can be seen that the full data estimator $\widehat r(x)$ has a non-negligible bias if $r^P(x)\neq r^Q(x)$. Asymptotically, the relative efficiency is defined by
$\mathbb{RE}[\widehat {r}_a^Q(x),\widehat r(x)]=\frac{\mathbb{MISE}_{\widehat r}}{\mathbb{V}_{\widehat {r}^Q_a}},$ where $\mathbb{V}_{\widehat {r}^Q_a}=\int_0^1\mathbb{V}[\widehat {r}_a^Q(x)]dx$ and $\mathbb{MISE}_{\widehat r}$ is the mean integrated square error of $\widehat r(x)$ defined by $\mathbb{MISE}_{\widehat r}=\int_0^1\mathbb{MSE}[\widehat r(x)]dx$.

\noindent{\bf Corollary 2.4.} {\it Under the conditions of Theorem 2.1, for the case of $0<\tau<\infty$, suppose that the full design dataset $D_X$ satisfies the quasi-uniform condition,  $n^P/n\rightarrow \tau_P\neq 0$ and $(n^Ph^P+n^Qh^Q)/(nh)\rightarrow\phi$, then

(i)
$\mathbb{RE}[\widehat {r}_a^Q(x),\widehat r(x)]\rightarrow\infty\mbox{ if } r^P(x)\neq r^Q(x)$  with $x\in (c_1,c_2)\subset [0,1]$  for some constants $c_2>c_1$;

(ii) $\mathbb{RE}[\widehat {r}_a^Q(x),\widehat r(x)]=\frac{\rho\phi}
{2+\tau+\frac{1}{\tau}}$ if P = Q. }

In the corollary, the quasi-uniform condition on the full design point set $D_X$ is defined as the same as in condition C2. Specifically, it means that there exists a common design function $\varphi(x)>0$ such that $s_i -s_{i-1}=1/(\varphi(X_i)n)+o(1/n)$ with $s_i=(X_i+X_{i+1})/2$ for all $X_i\in D_X$. We have the following observation from the corollary.

\noindent{\bf Remark 2.4.} {\it By the same argument as used in Remark 2.3, the bandwidth $h^Q$ is larger than the bandwidth $h$ in many cases. It shows that it is highly possible that $\phi>\frac{2+\tau+\frac{1}{\tau}}{\rho}$. In this case, the sw-TLE $\widehat {r}_a^Q$ is always more efficient than the full data estimator $\widehat r(x)$. Only for the case of P = Q, and $0\leq\phi<
\frac{2+\tau+\frac{1}{\tau}}
{\rho},$  the relative efficiency is low, i.e., $\mathbb{RE}[\widehat {r}_a^Q(x),\widehat r(x)]<1$.}

\subsubsection{Basis function representation}

Now we discuss the general framework of (\ref{(general)}). We use the general adjustment function in (\ref{(general)}) because $r^P(x)$ may be orthogonal to $r^Q(x)$ for some $x\in [0,1]$, and it will be shown in the Supplement Material that the ideal choice of $f(\widehat r^P(x),x)$ is an identity function, approximately, being smoother than $r^Q(x)$.  Note that a general function $f(\widehat r^P(x),x)$ can be approximated by basis functions. We thus consider the following representation: \begin{equation}\label{(polynomial)}f(\widehat {r}^P(x),x)=\xi_1(x)\phi_1(\widehat r^P(x))+\cdots+\xi_k(x)\phi_k(\widehat r^P(x)),\end{equation} where $\phi_j(x)$ are orthogonal basis functions, the coefficients $\xi_j(x)$ are unknown functions to be estimated, and the choice of the positive integer $k$ will be given after the condition C1' below. Denote $\bm{\xi}_k(x)=(\xi_1(x),\cdots,\xi_k(x))^T$, $\bm{\phi}_k^P(x)=(\phi_1({r}^P(x)),\cdots,\phi_k({r}^P(x)))^T,$  $\widehat{\bm\phi}_k^P(x)=(\phi_1(\widehat {r}^P(x)),\cdots,\phi_k(\widehat {r}^P(x)))^T$, and $
\dot{\bm\phi}_k^P(x)=(\dot\phi_1(r^P(x)),\cdots,\dot\phi_k(r^P(x)))^T$ with $\dot\phi_j(r^P(x))=\frac{d}{dx}\phi_j(r^P(x))$.
By minimizing the following local $L_2$-criterion
$$\int_0^1 K_{h^Q}(t-x)\{r^Q(t)-({\bm{\xi}}_k(x))^T\widehat{\bm\phi}_k^P(t)\}^2dt$$ for ${\bm{\xi}}_k(x)$, we get the solution of ${\bm{\xi}}_k(x)$ as
\begin{equation*}\begin{split}{\bm{\xi}}_{k,h^Q}(x)
=\left(\int_0^1K_{h^Q}(t-x)\widehat{\bm\phi}_k^P(t)(\widehat{\bm\phi}_k^P(t))^Tdt\right)^{-1}
\left(\int_0^1K_{h^Q}(t-x)r^Q(t)\widehat{\bm\phi}_k^P(t)dt\right).\end{split}\end{equation*}
Actually, the solution is a projection of each component of the vector of $\widehat {\bm\phi}_k^P(x)$ into the target function $r^Q(x)$. By the approximation to the integral as argument as in (\ref{integral}), we get the estimator of ${\bm{\xi}}_{k,h^Q}(x)$ as
\begin{equation*}\begin{split}\widehat{\bm{\xi}}_k(x)&=
\left(\int_0^1K_{h^Q}(t-x)\widehat{\bm\phi}_k^P(t)(\widehat{\bm\phi}_k^P(t))^Tdt\right)^{-1}
\left(\sum_{i=1}^{n^Q}Y_i^Q\int_{s_{i-1}^Q}^{s_i^Q}K_{h^Q}(t-x)
\widehat{\bm\phi}_k^P(t)dt\right),\end{split}\end{equation*}
where
$\widehat{\bm{\xi}}_k(x)=(\widehat\xi_1(x),\cdots,\widehat\xi_k(x))^T$.
Finally, we get the general sw-TLE of $r^Q(x)$ as
\begin{equation}\label{(polynomial-estimation)}
\widehat r^Q_{ak}(x)=(\widehat{\bm{\xi}}_k(x))^T\widehat{\bm\phi}_k^P(x).
\end{equation}

For the estimator of the basis function representation, the condition C1 is recast as
\begin{itemize}\item[C1'.] For all $x$, the functions
$\phi_j(r^P(x))$ and $r^Q(x)$
have the second-order continuous and bounded derivatives, $\|{\bm\phi}_k^P(x)\|\neq 0$, $\|{\bm\phi}_k^P(x)\|<c$ and $|(\bm{\xi}^0_k(x))^T\dot{\bm\phi}_k^P(x)|<c$ for a constant $c>0$, and $\langle \phi_j(r^P),r^Q \rangle_{h^Q}(x)\neq 0$ as $h^Q\rightarrow 0$ for some $j\in\{1,\cdots,k\}$, where $\bm{\xi}^0_k(x)=\frac{r^Q(x)}{\|\bm{\phi}_k^P(x)\|^2}
(\bm{\phi}_k^P(x))^T$.\end{itemize}
The condition C1' shows that we should choose the smallest positive integer $k$ such that the inner product $\langle \phi_j(r^P),r^Q \rangle_{h^Q}(x)\neq 0$ for all $x$.


\noindent{\bf Theorem 2.5.} {\it Under the fixed design model with the conditions C1', C2 and the regularity condition S1 and S2 in the Supplement Material, if the sample sizes and bandwidths satisfy the conditions given in Theorem 2.1, then the sw-TLE $\widehat r^Q_{ak}(x)$ in (\ref{(polynomial-estimation)}) has the following asymptotic normality:

(i) for $0<\tau<\infty$ and $ x\in(0,1)$,
\begin{equation*}\begin{split}&\sqrt{n^Ph^P+n^Qh^Q}
\left(\widehat {r}_{ak}^Q(x)-r^Q(x)\right)\\&\stackrel{d}\rightarrow
N\left(0,\left(1+\frac{1}{\tau}\right)\left(2\bm{\xi}^0_k(x))^T
\dot{\bm\phi}_k^P(x)\right)^2
\frac{\sigma^2_P}
{\varphi^P(x)}\int_{0}^{1}K^2(t)dt+\frac{\sigma^2_Q(1+\tau)
}
{\varphi^Q(x)}\int_{0}^{1}K^2(t)dt
\right);\end{split}\end{equation*}

(ii) for $\tau=\infty$ and $ x\in(0,1)$,
\begin{equation*}\begin{split}\sqrt{n^Qh^Q}
\left(\widehat {r}_{ak}^Q(x)-r^Q(x)\right)\stackrel{d}\rightarrow
N\left(0,\frac{\sigma^2_Q}
{\varphi^Q(x)}\int_{0}^{1}K^2(t)dt
\right);\end{split}\end{equation*}

(iii) for $\tau=0$ and $ x\in(0,1)$,
\begin{equation*}\begin{split}\sqrt{n^Ph^P}
\left(\widehat {r}_{ak}^Q(x)-r^Q(x)\right)\stackrel{d}\rightarrow
N\left(0,\left(2\bm{\xi}^0_k(x))^T
\dot{\bm\phi}_k^P(x)\right)^2
\frac{\sigma^2_P}
{\varphi^P(x)}\int_{0}^{1}K^2(t)dt\right).\end{split}\end{equation*}

}

By the theorem, we can get the conditions for a positive transfer learning, which are similar to those given in Corollary 2.2 and Corollary 2.3. For example, if $0<\tau<\infty$ and $k>1$, the condition C3 can be rewritten as $\rho>1$ and $\tau>\frac{\sigma_P^2\int_0^1\left(2\bm{\xi}^0_k(x))^T\dot{\bm\phi}_k^P(x)\right)^2\frac{1}{\varphi^P(x)}dx}{\sigma_Q^2\left (\rho-1\right)\int_0^1\frac{1}{\varphi^Q(x)}dx}.$ It shows again that the smoothness of  $\bm\xi_{k,h^Q}(x)$ and a relatively large value of $\tau$ can guarantee a positive transfer learning.
When $r^P(x)$ is orthogonal to $r^Q(x)$ for some $x\in [0,1]$, although the information of $r^P(x)$ cannot be transferred to $r^Q(x)$ by the projection, at least one basis function $\phi_j(r^P(x))$ contains the helpful information for inferring $r^Q(x)$ under the rule of the projection. Thus, the basis function-based sw-TLE can be applied to the case when $r^P(x)$ is orthogonal to $r^Q(x)$.
Similarly, we can compare the sw-TLE $\widehat {r}_{ak}^Q(x)$ with the full data G-M estimator $\widehat r(x)$. The relative efficiency is similar to that given in Corollary 2.4.

\subsection{The method for random design models}

Under random design model, as an example, the Nadaraya-Watson (N-W) estimator is used as a realization of $r^P(x)$, which is defined by
$\widehat {r}^P(x)=\frac{\sum_{i=1}^{n^P}Y_i^PK_{h^P}(X_i^P-x)}
{\sum_{i=1}^{n^P}K_{h^P}(X_i^P-x)}.$
In the following, we try to adjust the above estimator by the general form as in (\ref{(general)}) to approximate the target function $r^Q(x)$.

\subsubsection{Linear framework}

We first focus on the following linear form
$\widehat {r}^P(x)\eta(x),$
and employ the following local $L_2$-criterion
$\mathbb{E}\left[K_{h^Q}(X^Q-x)(Y^Q-\widehat {r}^P(X^Q)\eta(x))^2|D^P\right]$ to choose the adjustment factor $\eta(x)$ at a fixed target point $x$.
By minimizing the above criterion, we get the solution as
$\eta_{h^Q}(x)=
\frac{\mathbb{E}[K_{h^Q}(X^Q-x)Y^Q\widehat {r}^P(X^Q)|D^P]}{\mathbb{E}[K_{h^Q}(X^Q-x)
(\widehat {r}^P(X^Q))^2|D^P]}.$ It can be seen that a simple empirical version of the above solution can be expressed as
$\widehat\eta(x)=
\frac{\sum_{i=1}^{n^Q}K_{h^Q}(X_i^Q-x)Y_i^Q\widehat {r}^P(X_i^Q)}{\sum_{i=1}^{n^Q}K_{h^Q}(X_i^Q-x)
(\widehat {r}^P(X_i^Q))^2}.$
Consequently, based on the realization $\widehat {r}^P(x)$ from the P-model, the estimator $r^Q(x)$ for the Q-model can be chosen as
\begin{equation}\label{estimate-B} \widehat {r}_b^Q(x)=\widehat {r}^P(x)\widehat\eta(x).
\end{equation}
The structural reasonability of the estimator (\ref{estimate-B}) is the same as those of the estimator (\ref{estimate-A}) and the James-Stein estimator (see Subsection 2.1 and the Supplement Material). It is also a source-function weighted-transfer learning estimator (sw-TLE).
The CV criterion for the bandwidths is similar to (\ref{(CV)}). By minimizing the CV criterion, we can get the numerical solutions of the bandwidths.

For the sw-TLE (\ref{estimate-B}), the theoretical properties are given in the Supplement Material (see Theorem S.1, Corollary S.2 and Corollary S.3). The condition of Theorem S.1 shows that in addition to the possible dissimilarity between regression functions $r^P(x)$ and $r^Q(x)$ (conditional expectations drift), the density functions $\varphi^P(x)$ and $\varphi^Q(x)$ of $X^P$ and $X^Q$ may be dissimilar (cavariates shift).
Theorem S.1 indicates again that without the familiar similarity condition, for the case of $0<\tau<\infty$, it is possible that our sw-TLE can improve estimation efficiency.
The asymptotic normality given in Theorem S.1 seems to be the same as in the case of fixed design model. Although the asymptotic variances are the same, the asymptotic biases are actually
different. In other words, the asymptotic normality cannot embody the difference
between the neglected asymptotic biases in the condition of the under-smoothing.

For comparing the sw-TLE $\widehat {r}_b^Q(x)$ with the local data N-W estimator $\widehat r^Q(x)$ obtained only by the data from the target model Q,
the relative efficiency is given in Corollary S.2.
From the corollary, we can get the condition on the sample size of P-model for a positive (or negative) transfer learning; the detail is similar to that in Remark 2.3.
Finally, for comparing the sw-TLE $\widehat {r}_{b}^Q(x)$ with the full data N-W estimator $\widehat r(x)$, the relative efficiency is presented in Corollary S.3.
The corollary implies that  in most cases, the sw-TLE $\widehat {r}_{b}^Q(x)$ is more efficient than the full data estimator $\widehat r(x)$.

\subsubsection{Basis function representation}

Note that $r^P(x)$ and $r^Q(x)$ are possible to be orthogonal to each other for some $x\in [L,U]$. Hence, it is necessary to extend the above to a general case.
Similar to (\ref{(polynomial)}), we consider the case where the function $f(\cdot,\cdot)$ under the random design model is chosen as the form: $f(\widehat {r}^P(x),x)=\eta_1(x)\phi_1(\widehat {r}^P(x))+\cdots+\eta_k(x)\phi_k(\widehat {r}^P(x)),$ where the coefficients $\eta_j(x)$ are unknown functions to be determined. As shown in the previous subsection, we should the smallest positive integer $k$ such that the inner product $\langle \phi_k(r^P),r^Q \rangle_{h^Q}(x)\neq 0$ for all $x$. Denote $\bm{\eta}_k(x)=(\eta_1(x),\cdots,\eta_k(x))^T$ and $\widehat{\bm{\phi}}^P_k(x)=(\phi_1(\widehat {r}^P(x)),\cdots,\phi_k(\widehat {r}^P(x)))^T$.
By the same argument as used above, we get the estimator of $\bm{\eta}_k(x)$ as
\begin{equation*}\begin{split}\widehat{\bm{\eta}}_k(x)&=
\left(\sum_{i=1}^{n^Q}K_{h^Q}(X^Q_i-x)\widehat{\bm{\phi}}^P_k(X_i^Q)(\widehat{\bm{\phi}}^P_k(X_i^Q))^T\right)^{-1}
\left(\sum_{i=1}^{n^Q}K_{h^Q}(X^Q_i-x)Y_i^Q
\widehat{\bm{\phi}}^P_k(X_i^Q)\right).\end{split}\end{equation*}
Finally, for random design, we attain the sw-TLE of $r^Q(x)$ as
\begin{equation}\label{estimate-B-1}\widehat r^Q_{bk}(x)=(\widehat{\bm{\eta}}_k(x))^T\widehat{\bm{\phi}}^P_k(x).\end{equation}

The theoretical property of the sw-TLE $\widehat{\bm{\eta}}_k(x)$ is reported in Theorem S.4 in the Supplement Material.
Generally, similar to Corollary 2.2 and Corollary 2.3, a relatively large sample size of the P-model is benefit for the enhancement of estimation efficiency.

\setcounter{equation}{0}
\section{Extensions}

In this section, we first consider a nonparametric target model combined with a single parametric source. The details for the extension to the case with multi-source will be presented in the Supplement Material.

The methodology proposed in the previous section can be extended into the following semiparametric models:
\begin{equation*} \begin{split} \mbox{(P-model)} \ Y_i^P=r^P(X_i^P,\theta)+
\varepsilon_i^P; \ \
  \mbox{(Q-model)}  \ Y_i^Q=r^Q(X_i^Q)+\varepsilon_i^Q,\end{split}\end{equation*} where the error terms satisfy the same conditions as in (\ref{model-P}).
In the above models, $r^P(x,\theta)$ is a known function up to an unknown parameter $\theta\in\Theta\subset R^d$, and $r^Q(x)$ is an unknown nonparametric function. Here we particularly assume that $r^P(x,\theta)$ is a nonlinear function of $\theta$. It is because if  $r^P(x,\theta)$ is linear in $\theta$, the resulting sw-TLE  (\ref{estimate-C}) is free of $\theta$. Our purpose is to estimate the nonparametric target function $r^Q(x)$. We only consider the case where both P-model and Q-model are random design models with $x\in[L,U]$. For fixed design models, the method is similar.
Under the above models, the seeming similarity condition C1 is redefined by
\begin{itemize}
\item[C1''.] For $x\in[L,U]$,
$r^P(x,\theta)\neq 0$, and the inner product $\langle r^Q(X^Q),r^P(X^Q,\theta)\rangle_{h^Q}(x)\neq 0$ as $h^Q\rightarrow 0$, where $\theta$ is in an neighbourhood of its true value $\theta^0$.\end{itemize}
Let $\widehat\theta$ be a $\sqrt{n^P}$-consistent estimator of $\theta$, for example, the least squares (LS) estimator obtained  by the data from P-model. To approximate the target function $r^Q(x)$, we adjust the estimator $r^P(x,\widehat\theta)$ to the form:
$r^P(x,\widehat\theta)\alpha(x),$
and employ the local $L_2$-criterion
$\mathbb{E}\left[K_{h^Q}(X^Q-x)(Y^Q-
r^P(X^Q,\widehat\theta)\alpha(x))^2|D^P\right]$ to choose $\alpha(x)$. By minimizing the above criterion, we get the solution as
$\alpha_{h^Q}(x)=
\frac{\mathbb{E}[K_{h^Q}(X^Q-x)Y^Qr^P(X^Q,\widehat\theta)|D^P]}{\mathbb{E}[K_{h^Q}(X^Q-x)
(r^P(X^Q,\widehat\theta))^2|D^P]},$ and its empirical version as
$\widehat\alpha(x)=
\frac{\sum_{i=1}^{n^Q}K_{h^Q}(X_i^Q-x)Y_i^Qr^P(X_i^Q,\widehat\theta)}{\sum_{i=1}^{n^Q}K_{h^Q}(X_i^Q-x)
(r^P(X_i^Q,\widehat\theta))^2}.$
Finally, we attain the sw-TLE of the target function $r^Q(x)$ as
\begin{equation}\label{estimate-C}\widehat {r}_c^Q(x)=r^P(x,\widehat\theta)\widehat\alpha(x).\end{equation} Actually, the above is a semiparametric estimator, starting out with a parametric approximation, and ending up with a nonparametric estimation. For the estimator above, the CV criterion for choosing bandwidth is similar to (\ref{(CV)}).

\noindent{\bf Theorem 3.1.} {\it Under the semiparametric models above and the regularity conditions S5-S9 given in the Supplement Material, if $r^P(x,\theta)$ is a nonlinear function of $\theta$, $r^P(x,\theta)$ and $r^Q(x)$ satisfy the seeming similarity condition C1'', and the sample sizes and bandwidth satisfy
$h^Q=o((n^Q)^{-1/5})$, and $n^P/(n^Qh^Q)\rightarrow \tau$, then the sw-TLE (\ref{estimate-C}) has the following asymptotic normality:

(i) for $0<\tau<\infty$ and $x\in(L, U)$,
\begin{equation*}\begin{split}& \sqrt{n^P+n^Qh^Q}
\left(\widehat {r}_c^Q(x)-r^Q(x)\right)\\&\stackrel{d}\rightarrow
N\left(0,\sigma^2_P\left(1+\frac{1}{\tau}\right)\phi(\theta^0)
\left(\frac{2r^Q(x)
}{r^P(x,\theta^0)}\right)^2+\frac{\sigma^2_Q(\tau+1)}
{\varphi^Q(x)}\int_{L}^{U}K^2(t)dt
\right);\end{split}\end{equation*}

(ii) for $ \tau=\infty$ and $x\in(L, U)$,
\begin{equation*}\begin{split}  \sqrt{n^Qh^Q}
\left(\widehat {r}_c^Q(x)-r^Q(x)\right) \stackrel{d}\rightarrow
N\left(0,\frac{\sigma^2_Q }
{\varphi^Q(x)}\int_{L}^{U}K^2(t)dt
\right);\end{split}\end{equation*}

(iii) for $\tau=0$ and $x\in(L, U)$,
\begin{equation*}\begin{split} \sqrt{n^P}
\left(\widehat {r}_c^Q(x)-r^Q(x)\right) \stackrel{d}\rightarrow
N\left(0,\sigma^2_P \phi(\theta^0)
\left(\frac{2r^Q(x)
}{r^P(x,\theta^0)}\right)^2\right),\end{split}\end{equation*}
where
$\phi(\theta)=(\mathbb{E}[\dot r^P(X^P,\theta)])^T
(\mathbb{V}_r^P(\theta))^{-1}\mathbb{E}[\dot r^P(X^P,\theta)]
$ and $\mathbb{V}_r^P(\theta)$ is defined in S9  in the Supplement Material.
}


The theorem ensures that for the case $0<\tau<\infty$, it is possible that the estimator can enhance estimation efficiency.
Furthermore, we can compare the sw-TLE $\widehat {r}_c^Q(x)$ with the local data N-W estimator $\widehat r^Q(x)$ obtained only by the data of the target model Q. When constructing relative efficiency for the two estimators, the asymptotic biases can be ignored relative to the asymptotic variance due to under-smoothing.
The following corollary presents the properties of the relative efficiency.

\noindent{\bf Corollary 3.2.} {\it Under the conditions of Theorem 3.1, for the case of $0<\tau<\infty$, if $h^Q/\widetilde h^Q\rightarrow \rho$, then, asymptotically, the relative efficiency can be expressed as
\begin{equation}\label{(efficiency-1)}\mathbb{RE}[\widehat {r}_c^Q(x),\widehat r^Q(x)]=\frac{\rho\tau\int_L^U
\frac{1}{\varphi^Q(x)}dx\int_{L}^{U}K^2(x)dx}{\sigma^2_P\phi(\theta^0)
\int_L^U\left(2\frac{r^Q(x)
}{r^P(x,\theta^0)}\right)^2dx+\tau\sigma^2_Q\int_L^U \frac{1}
{\varphi^Q(x)}dx\int_{L}^{U}K^2(x)dx}.\end{equation}
Consequently, we have $\mathbb{RE}[\widehat {r}_c^Q(x),\widehat r^Q(x)]> 1$ if and only if the following condition holds:
\begin{itemize}
\item[C4.] $\rho>1$ and $\tau>\frac{\sigma^2_P\phi(\theta^0)\int_L^U\left(2\frac{r^Q(x)
}{r^P(x,\theta^0)}\right)^2dx}{\sigma^2_Q \left(\rho-1\right)\int_L^U
\frac{1}{\varphi^Q(x)}dx\int_{L}^{U}K^2(x)dx}.$
\end{itemize}
}

Generally, the method proposed can be extended to  multi-source case.
As an example, we consider
nonparametric target model with the multi-source parametric models:
\begin{equation*}\mbox{(P$_j$-model)} \ \ \ Y_{ij}^P=r_j^P(X_{ij}^P,\theta_j)+\varepsilon_{ij}^P,i=1,
\cdots,n_j^P,j=1,\cdots,m,\end{equation*} where the errors satisfy $\mathbb{E}[\varepsilon_{ij}^P|X_{ij}^P]=0$ and $\mathbb{V}[\varepsilon_{ij}^P|X_{ij}^P]=\sigma^2_{Pj}$. The details about the estimation method and theoretical property are given in the Supplement Material.

\setcounter{equation}{0}
\section{Numerical studies}
We use simulation studies and real data example to show the performance of the proposed sw-TLE. Due to the length limit, only the main results of the similar source problem and the real data analysis are reported here, some additional results for the similar source problem are listed in the Supplement Material. Furthermore, more numerical studies for identical source, unrelated source and multi-source problems can be found in the Supplement Material as well.
\subsection{Elementary empirical evidences}\label{sect:simulation}
\begin{figure*}[t]
\centering
\includegraphics[width=0.8\textwidth]{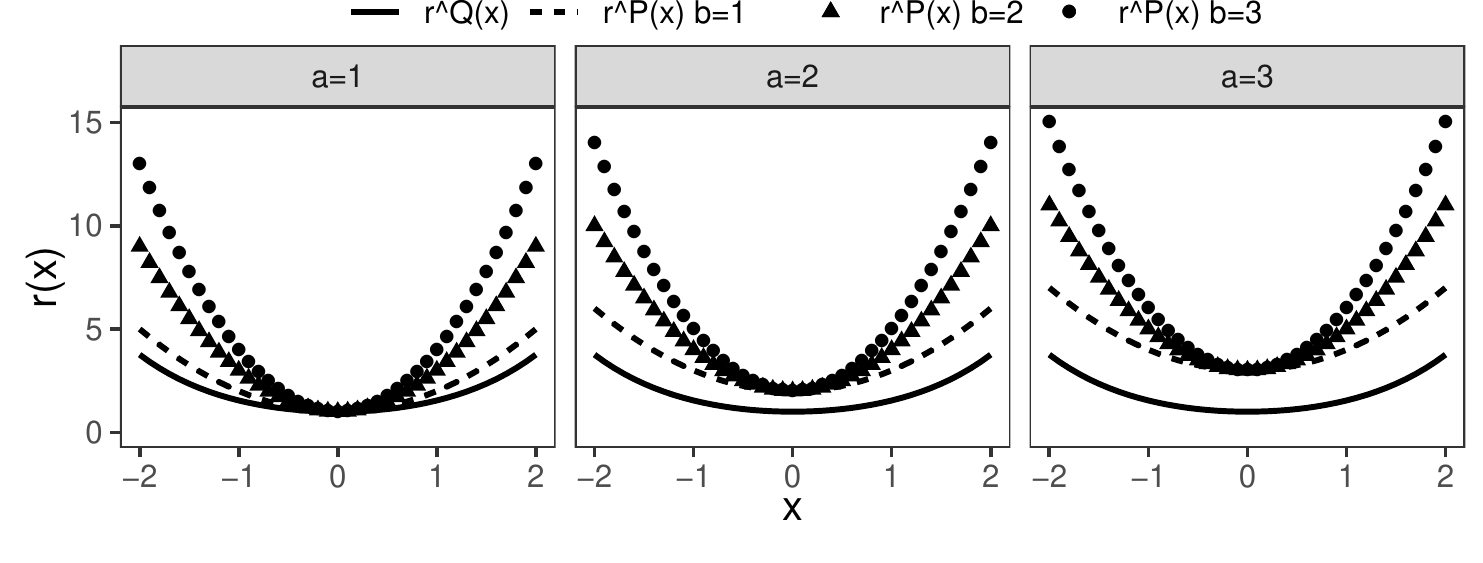}
\caption{\small The divergence between source and target with $r^P(x)=a+b x^2$ and $r^Q(x)=\cosh(x)$ . } \label{fig:example}
\end{figure*}

We first consider the following models:
$$
  \text{Model P}: Y^P=a+b (X^P)^2+\varepsilon^P   \text{ and }
  \text{Model Q}: Y^Q=\cosh(X^Q)+\varepsilon^Q,
$$
where independent random variables $X^P,X^Q\sim U[-2,2]$, and $\varepsilon^P,\varepsilon^Q\sim N(0,0.2^2)$. Here, the constants $a$ and $b$ valued in $\{1, 2, 3\}$ are employed to determine the level of divergence between the source and the target, see Fig. \ref{fig:example}.

In the simulation, we consider separately two cases: $n^P=500$ is fixed with varying $n^Q$, and $n^Q=50$ is fixed with varying $n^P$  to show the effect of the data size. Our sw-TLE is compared with other three methods:
\begin{itemize}
  \item[1)] The Q-NW: the N-W regression with the data only from target model Q.
  \item[2)] The SA: the simple average of the N-W estimators, defined by 
  $$
  \widehat{r}^Q_{SA}(x)=\frac{\sqrt{n^Ph^P}\widehat{r}^P_{nw}(x)
  +\sqrt{n^Qh^Q}\widehat{r}^Q_{nw}(x)}{\sqrt{n^Ph^P}+\sqrt{n^Qh^Q}},
  $$
  where $\widehat{r}^P_{nw}(x)$ and $\widehat{r}^Q_{nw}(x)$ are the N-W estimators for P-model and Q-model, respectively.
  \item[3)] The WA: the data-driven weighted average of the N-W estimators, defined by 
  $
  \widehat{r}^Q_{WA}(x)=w^P\widehat{r}^P_{nw}(x)+w^Q\widehat{r}^Q_{nw}(x),
  $
  where $w^P$ and $w^Q$ are chosen by the criterion
  $
    (w^P, w^Q)^\top={\arg\min}_{w^P+w^Q=1 \atop w^P, w^Q \in [0,1]}\sum_{i=1}^{n^Q}(Y^Q_i-\widehat{r}^Q_{WA(-i)}(X_i^Q))^2
  $
  with $\widehat{r}^Q_{WA(-i)}(X_i^Q)$ being the leave-one-out form of WA.
\end{itemize}
The estimation performance is measured with the mean integrated squared error (MISE) derived by 1000 replications. All the kernel estimators are constructed by the Gaussian kernel and the bandwidths are chosen by the CV criterion given in (\ref{(CV)}).
The MISE curves are reported in Fig. \ref{fig:e2_np} and \ref{fig:e2_nq}, and more details can be found in Table S.4 and Table S.5 in the Supplement Material. We have the following findings:
\begin{enumerate}
  \item The sw-TLE is much better than the Q-NW, the SA and the WA under various choices of $\{a,b\}$, see Fig. \ref{fig:e2_np} and \ref{fig:e2_nq}. The SA even has over-range MISE curves,  and most time its MISE is larger than 5, see Table S.4 and Table S.5 in the Supplement Material.
  \item Unlike the effect on the three competitors, the divergence between P-model and Q-model has only a weak influence on the behavior of the sw-TLE. 
  \item The convergence rate of the sw-TLE is faster than the others.
  \item The sample size $n^P$ has a limited effect on the performance of the sw-TLE; see Table S.5 in the Supplement Material. The observation is consistent with Remark 2.1 (iii).
\end{enumerate}
\begin{figure*}[t]
\centering
\includegraphics[width=0.9\textwidth]{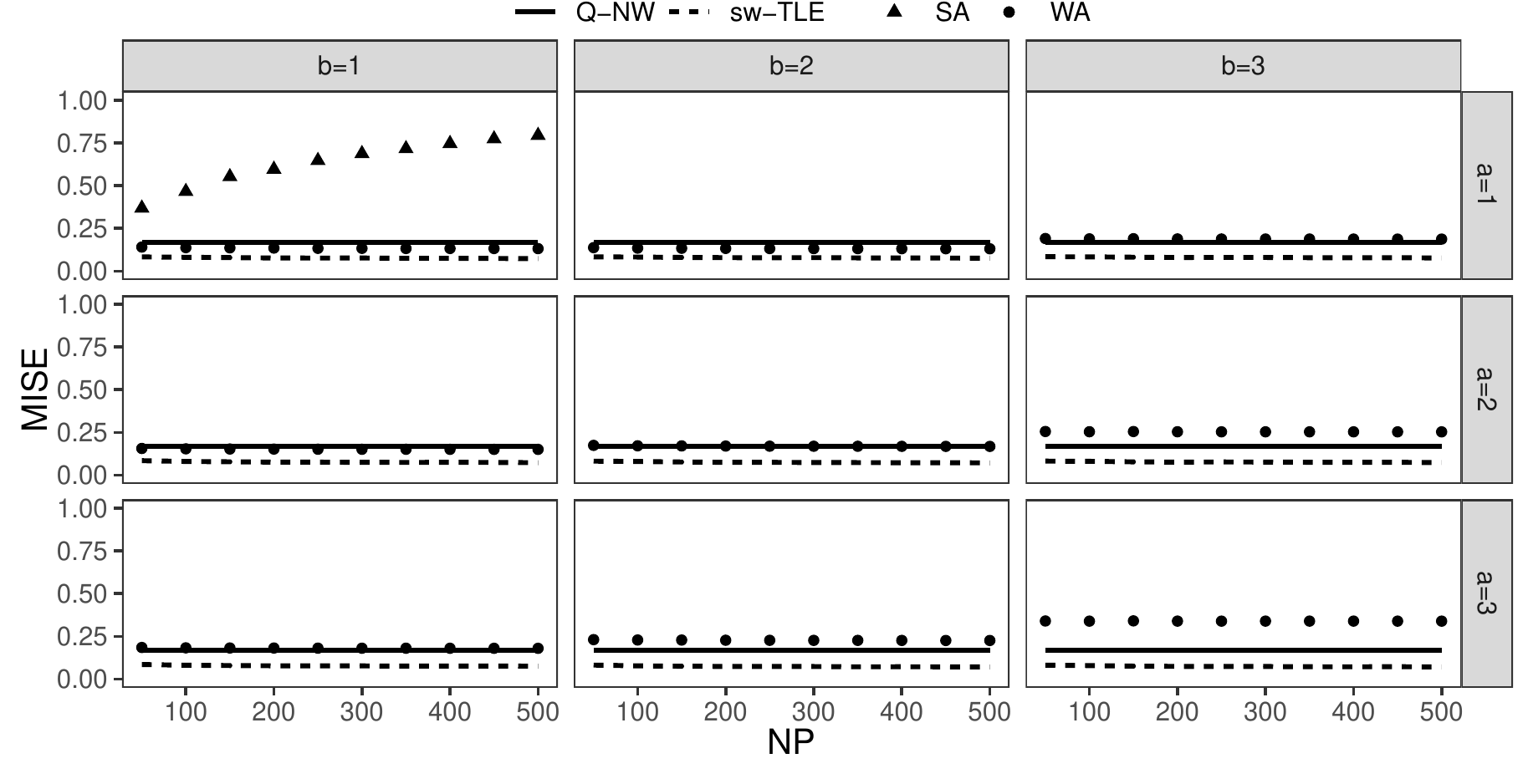}\\
\caption{\small The MISE curves of the estimators in the example with fixed $n^Q=50$. Note that the curves of MISE of the SA are beyond the realm of the figure except for the case of $a=1$ and $b=1$. }
\label{fig:e2_np}
\end{figure*}

\begin{figure*}[t]
\centering
\includegraphics[width=0.9\textwidth]{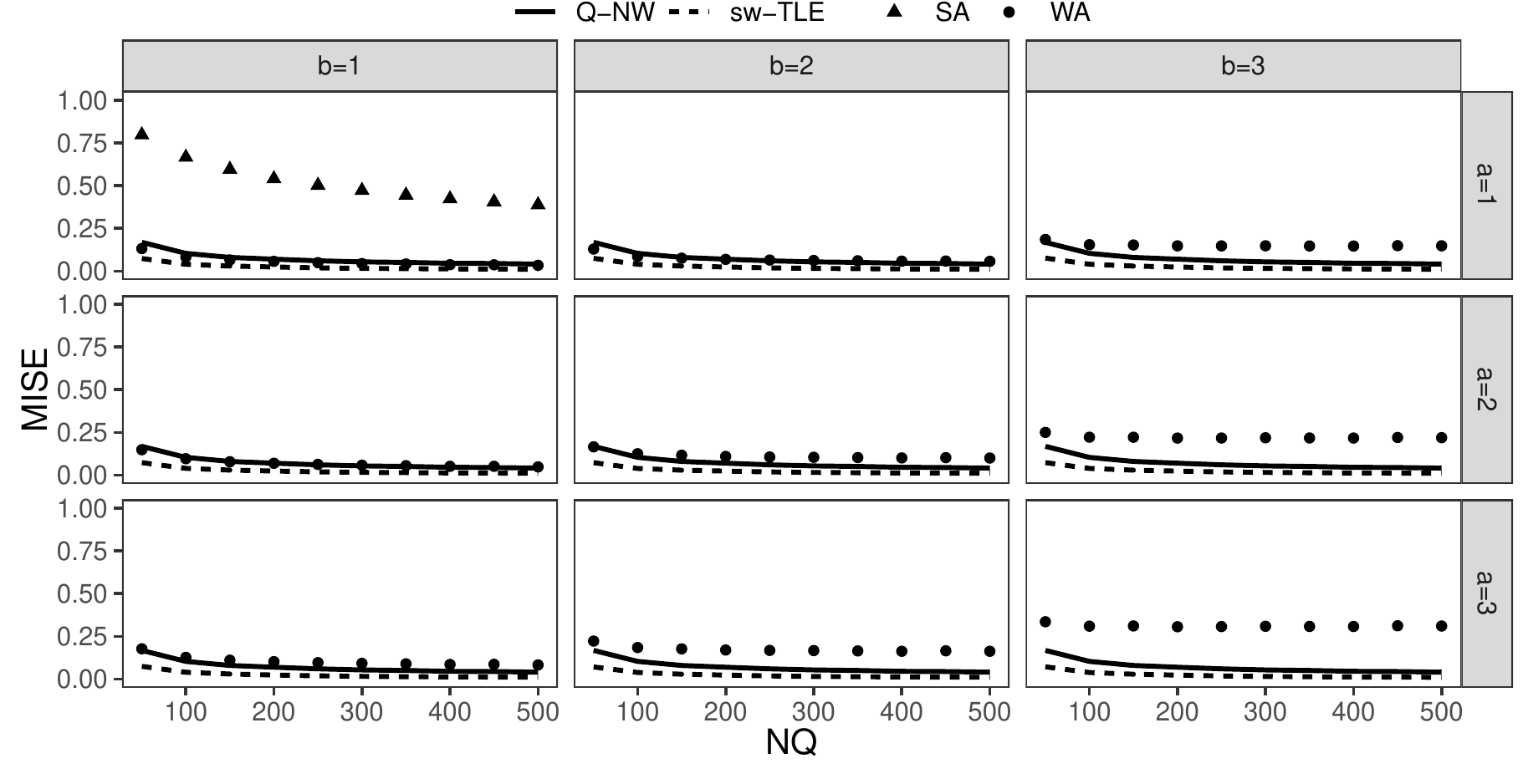}\\
\caption{\small The MISE curves of the estimators in the example with fixed $n^P=500$. Note that the curves of MISE of the SA are beyond the realm of the figure except for the case of $a=1$ and $b=1$. }
\label{fig:e2_nq}
\end{figure*}
\subsection{Further simulation results}
We also investigate the following three situations: i) the identical source problem; ii) the unrelated source problem; iii) the multi-source problem. Only the main conclusions are summarized here, and more details are presented in the Supplement Material. We have the following observations:
\begin{enumerate}
  \item Under all the situations, the sw-TLE is much better than the SA and the WA.
  \item In the case of identical source,  our method is even better than the full data N-W regression when $n^P/n^Q$ are not too large. This conclusion is indicated in Remark 2.4.
  \item Our method has good adaptability to the case with unrelated source. When $n^P/n^Q$ is not too large, the sw-TLE is better than Q-NW. But the performance of the sw-TLE will become bad when $n^P/n^Q$ is too large, which is consistent with conclusion in Remark 2.1 (iii).
  \item In the multi-source problem, the difference between sizes of two source data is influential in the performance of the sw-TLE.
\end{enumerate}

In short, all the simulation results (including those in the Supplement Material) can clearly verify all the theoretical conclusions given in the previous sections.

\subsection{Real data application}
The water-cement ratio of the fresh concrete mix is one of the main factors determining the quality and properties of hardened concrete. Now we want to investigate the influence of the water-cement  ratio on the SLUMP, the FLOW and the 28-day Compressive Strength by the Concrete Slump Test Data Set \footnote{http://archive.ics.uci.edu/ml/datasets/Concrete+Slump+Test} in  \citet{yeh2007modeling}. The initial dataset included only 78 data and the other 25 data points were gotten after several years. So the set of the first 78 data points is treated as training set, while the rest is regarded as the test set. In our analysis, the covariate is set to be the water-cement ratio, while the SLUMP, the FLOW and the 28-day Compressive Strength (CS) are separately considered as the response variables. For the insufficiency of data, we utilize the sw-TLE method separately with the real source data (denoted as R-sw-TLE) and  the artificial source data (denoted as A-sw-TLE) to improve the effectiveness of modeling.

The real source data with size 424 consists of instances for age $=28$  in Concrete Compressive Strength Data Set\footnote{http://archive.ics.uci.edu/ml/datasets/Concrete+Compressive+Strength} \citep{yeh1998modeling}. The dataset includes the information on cement, water and compressive strength of the concrete. Despite it has enough data size,  there is no information of the SLUMP and the FLOW in it. The real source model is similar to the target model about the 28-day Compressive Strength and, but is far from the target models separately with the FLOW and the SLUMP as the responses, see Fig. S.5 in the Supplement Material. 

The artificial covariate in source data set is chosen as 
$X \sim U[0.48,1.79]$, and the response variables are set separately as normal distributed random variables with mean and variance as those of the corresponding target data. The size of the artificial source data set is 200.

\begin{table}[t]
\renewcommand\arraystretch{0.7}
\caption{\label{tab:real} \small The MSRR and MSPE in real data analysis.}
\centering
\tiny
\begin{tabular}{ccccccc }
\hline
\multirow{2}{*}{}&\multicolumn{3}{c}{MSRR}&\multicolumn{3}{c}{MSPE}\\
\cline{2-7}
&R-sw-TLE&A-sw-TLE& Q-NW&R-sw-TLE&A-sw-TLE& Q-NW\\
\hline
CS&34.2447&32.9537 &44.7861 &44.5395&45.3495 &46.5829 \\
FLOW&170.3446&157.7408 &213.6381 &382.2413&376.0140 &440.9734 \\
SLUMP&48.7528&44.7528 &58.1126 &110.5239&108.6473 &113.8142 \\
\hline
\end{tabular}
\end{table}

\begin{figure*}[t]
\centering
\includegraphics[width=0.8\textwidth]{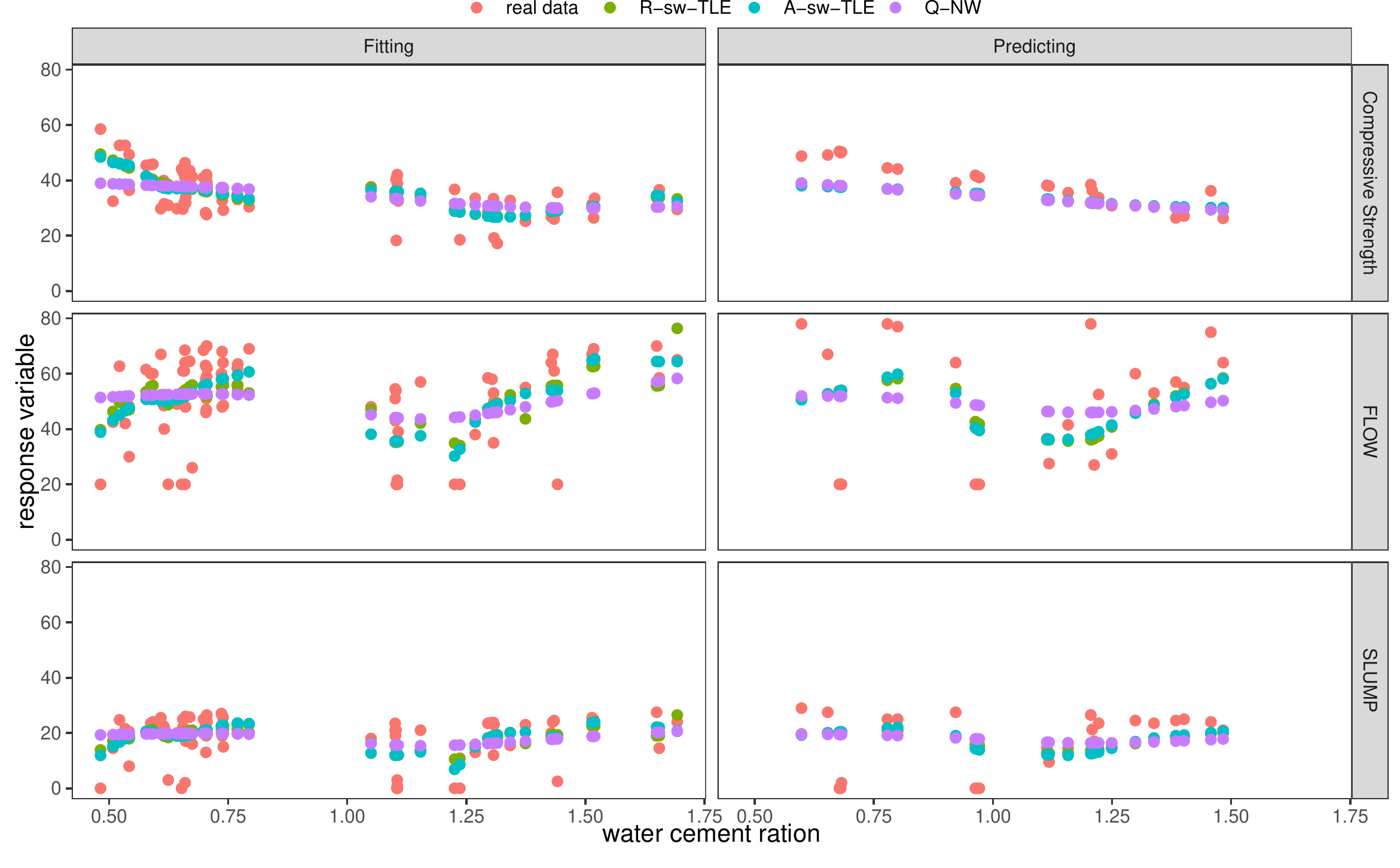}
\caption{\small The real data analysis }
\label{fig:real_real}
\end{figure*}

Our method is compared with the N-W estimation obtained only by the target data. Table \ref{tab:real} reports the mean squared regression residual (MSRR) and  mean squared prediction error (MSPE) defined as
$
\frac{1}{\#\{A\}}\sum_{i\in A}(Y^Q_i-\widehat{Y}^Q_i)^2,
$
where $\#\{A\}$ is the cardinality of $A=\{1, \cdots, 78\}$ for the MSRR and $A=\{79, \cdots, 103\}$ for the MSPE.  Fig. \ref{fig:real_real} shows the regression and prediction curves. We have the following findings:
\begin{enumerate}
  \item There exist nonlinear relations separately between the water-cement ratio and the 28-day Compressive Strength, the FLOW and the SLUMP. These are consistent with the existing knowledge of the cement.
  \item Our method is better than the N-W estimation both in the fitting and  predicting in the sense that the regression residuals and prediction errors of our method are less than those of the N-W estimation.
  \item Although the gap is narrow, the results gotten from artificial source datasets are better than those from the real source data; see Table \ref{tab:real} and Fig. \ref{fig:real_real}. It means that sometimes our method can overcome the obstacle of finding source data, a common  difficulty in transfer learning.
\end{enumerate}

The real data analysis can further illustrate the theoretical conclusions obtained.

\setcounter{equation}{0}
\section{Conclusions and future works}

As stated in the previous sections, the similarity condition seems to be essential to efficient transfer learning, the existing transfer learning methodologies are based on the belief that the one can transfer knowledge across the tasks by discovering the similar characteristics among the related domains.
However, the similarity condition is usually difficult to check or even is violated in practice. For this reason, instead of the similarity condition, a seeming similarity condition was introduced in the previous sections. The seeming similarity is based on a non-orthogonality together with a smoothness of the adjustment functions, and is naturally satisfied under common situations and even could be a dissimilarity condition in some sense.
Under the
seeming similarity condition, an $L_2$-adjustment strategy was proposed and then a source-function weighted-transfer learning was obtained in the previous sections. By source-function weighting, the variance of the proposed estimator can be significantly reduced. Consequently, the new strategy can achieve the global convergence rate and enhance the estimation efficiency. Thus, the new transfer learning has a distinct advantage over the classical methods for the case when the sample size of the source models is large, and the familiar similarity condition is violated. Moreover, the strategy does apply to nonparametric and semiparametric models.
The behavior of the transfer learning method was further illustrated by various numerical examples from simulation experiments and a real data analysis. The numerical examples further verified that the
finite performance of the new method is much better than the competitors. It is somewhat surprising that the theoretical properties and numerical results together indicated that even for the case where the source models are unrelated to the target model, the information of source models can be used to improve the transfer learning estimation. The new theory was further supported by more analysises such as the smoothness of the adjustment function and the structural reasonability given in the Supplement Material.

In the previous sections, we only focused on the estimation of regression functions. All the estimators of the regression functions have closed-form expressions. With the closed-form expressions, we can adjust the estimators from the source functions to the target function. In most situations, however, the estimators are not linear functions of the response variables, and moreover,
often have no closed-form expression. In these complicated situations, the solution only can be obtained numerically by iterative algorithms, such as the Newton-Raphson algorithm. Thus, it is still a challenge to extend the strategy to these complex cases. Furthermore, in this paper, the $L_2$-adjustment is applied to function estimation, rather than parameter estimation. Thus, it is desired to extend the method to parameter estimation. Theoretically,  it is still an open question what kind of knowledge we gain from some source domains can be transferred into a target domain to improve the inference on the target model without the similarity condition.
These are
interesting issues and are worth further study in the future.
\bibliographystyle{Chicago}
\bibliography{ref}

\newpage
\bigskip
\bigskip
\begin{center}
\LARGE{\textbf{Supplement Material}}
\end{center}

\setcounter{equation}{0}
\setcounter{section}{0}
\section{Appendix}

\subsection{Smoothness of adjustment function}

In our method, the smoothness of the adjustment function is a key for a positive transfer learning.
We first check the smoothness of $\xi_{h^Q}(x)$ defined in (2.5).
By Lemma S.3 given below, we have
\begin{equation*}\begin{split}\xi_{h^Q}(x)=\frac{r^Q(x)}{{r}^P(x)}
+O_p((h^P)^2+1/\sqrt{n^Ph^P}).\end{split}\end{equation*}
It shows that we only need to compare the smoothness of the ratio function $\xi_o(x)=\frac{r^Q(x)}{{r}^P(x)}$ and the target function $r^Q(x)$. It is known that the integral of the second-order derivative squared, for example $\int_0^1(\ddot \xi_o(x))^2dx$, is commonly used to measure the smoothness. In Remark 2.3,  this criterion was also utilized  to compare the smoothness of $\xi_o(x)$ and $r^Q(x)$, and then to select the bandwidths.
Note that
if $\dot \xi_o(x)$ is a monotone function, we have $$\int_{0}^{1}(\ddot \xi_o(x))^2dx=\int_{\dot \xi_o(0)}^{\dot \xi_o(1)}\ddot \xi_o(x)d\dot \xi_o(x).$$ By the equivalence, the integrals $\int_{\dot \xi_o(0)}^{\dot \xi_o(1)}\ddot \xi_o(x)d\dot \xi_o(x)$ and $\int_{\dot r^Q(0)}^{\dot r^Q(1)}\ddot r^Q(x)d\dot r^Q(x)$ are employed to measure the smoothness of $\xi_o(x)$ and $r^Q(x)$.

We first consider the case of $\ddot \xi_o(x)> 0$ and $\ddot r^Q(x)>0$ for all $x\in[0,1]$. Suppose that the relationship between $r^P(x)$ and $r^Q(x)$ can be expressed as $r^P(x)=g(r^Q(x))$ for a smooth function $g(u)$, without loss of generality. Denote by $\dot g(u)$ the derivative of $g(u)$.
Then,
$$\dot \xi_o(x)
=\frac{\dot r^Q(x)(g(r^Q(x))-\dot g(r^Q(x)) r^Q(x))}{(g(r^Q(x)))^2}.$$  Denote $a_1=\dot \xi_o(0)$, $b_1=\dot \xi_o(1)$, $a_2=\dot r^Q(0)$ and $b_2=\dot r^Q(1)$. Consequently,
\begin{equation*}\begin{split}\int_{\dot \xi_o(0)}^{\dot \xi_o(1)}|\ddot \xi_o(x)|d\dot \xi_o(x)&=\dot \xi_o(b_1)-\dot \xi_o(a_1)\\&=\frac{\dot r^Q(b_1)(g(r^Q(b_1))-\dot g(r^Q(b_1)) r^Q(b_1))}{(g(r^Q(b_1)))^2}\\&\ \ \ -\frac{\dot r^Q(a_1)(g(r^Q(a_1))
-\dot g(r^Q(a_1)) r^Q(a_1))}{(g(r^Q(a_1)))^2}.\end{split}\end{equation*} Similarly,
$\int_{\dot r^Q(0)}^{\dot r^Q(1)}|\ddot r^Q(x)|d\dot r^Q(x)=\dot r^Q(b_2)-\dot r^Q(a_2)$.
We then introduce the following criterion for describing that the $\xi_{h^Q}(x)$ is smoother than $r^Q(x)$:
\begin{itemize}\item[S0.]
$0<c_1\dot r^Q(b_1)-c_0\dot r^Q(a_1)<\dot r^Q(b_2)-\dot r^Q(a_2),$ where
\begin{equation*}\begin{split}c_1=\frac{g(r^Q(b_1))-\dot g(r^Q(b_1)) r^Q(b_1)}{(g(r^Q(b_1)))^2}\mbox{ and } c_0=\frac{g(r^Q(a_1))-\dot g(r^Q(a_1)) r^Q(a_1)}{(g(r^Q(a_1))^2}.\end{split}\end{equation*}
\end{itemize}

1) Consider the special case of the linear framework: $$g( r^Q(x))=d r^Q(x)+e$$ for some constants $d$ and $e$. Under the framework, we have
\begin{equation*}\begin{split}c_1=
\frac{e}{(d r^Q(b_1)+e)^2}\mbox{ and } c_0=\frac{e}{(d r^Q(a_1)+e)^2}.\end{split}\end{equation*}
Particularly, it is possible that $ r^P(x)$ captures the
main features of the shape of the target regression function $ r^Q(x)$, which means the value of $|e|$ is small, and $d\neq0$. Note that  $\ddot \xi_o(x)> 0$ and $\ddot  r^Q(x)>0$  imply $\dot  r^Q(b_1)>\dot  r^Q(a_1)$ and $\dot  r^Q(b_2)>\dot  r^Q(a_2)$. Hence, under the above framework, when  $|e|$ is small enough, the smoothness condition S0 is satisfied naturally.
For example, consider the following condition:
\begin{itemize}\item[S0'.]
$|e|<\min\left\{c_2|d r^Q(a_1)|,c_2|d r^Q(b_1)|,\frac12\frac{c_3^2(1-c_2)^2}{\dot r_M}(\dot  r^Q(b_2)-\dot r^Q(a_2))\right\}$ for some constants $0<c_2<1$ and $0<c_3<\min\{
|d r^Q(a_1)|,|d r^Q(b_1)|\}$, where $\dot r_M=\max\{|\dot r^Q(b_1)|,|\dot  r^Q(a_1)|\}$.
\end{itemize}
The condition S0' means that $|e|$  is small compared to $|d|$,  $|r^Q(a_1)|$ and $| r^Q(b_1)|$.
It can be verified that the condition S0' implies the smoothness condition S0. These ensure that under linear framework as well as approximate similarity condition, the smoothness condition S0 holds.

2) Consider a class of nonlinear relationships between $r^P(x)$ and $r^Q(x)$ as
$$r^P(x)=g(r^Q(x))=F^{-1}\left(\log|r^Q(x)|+c\right) \mbox{ for some constant } c,$$
where $F(u)=\int^{u}_{0}\frac{1}{t+f(t)}dt$, and $f(u)$ satisfies $f(u)>0$ and $f(u)+u\neq 0$. Because of the option attribute of $f(u)$ and $c$, the above is a broad class of functions. It can be seen that the above relationship is the solution to the following differential equation:
$$g(r^Q(x))-\dot g(r^Q(x))r^Q(x)=-f(g(r^Q(x)).$$
Thus, we have
\begin{equation*}\begin{split}c_1=\frac{-f(g(r^Q(b_1))}{(g( r^Q(b_1)))^2}\mbox{ and } c_0=\frac{-f(g(r^Q(a_1))}{(g( r^Q(a_1)))^2}.\end{split}\end{equation*} Consequently,
when function $f(u)$ further satisfies that $|f(g(r^Q(a_1))|$ and
$|f(g(r^Q(b_1))|$ are small compared to $|g( r^Q(a_1))|$ and $|g( r^Q(b_1))|$, the smoothness condition S0 holds.

3) Generally, it can be verified that the the smoothness condition S0 holds under one of the following conditions:

(i) For arbitrary $\dot r^Q(a_1)$, $\dot r^Q(b_1)$, $\dot r^Q(a_2)$ and $\dot r^Q(b_2)$, $c_0$ and $c_1$ satisfy $0<c_0=c_1<\frac{\dot r^Q(b_2)-\dot r^Q(a_2)}{\dot r^Q(b_1)-\dot r^Q(a_1)}$;

(ii) For $\dot r^Q(b_1)>0$, $c_0$ and $c_1$ satisfy
$c_1<\frac{\dot r^Q(a_1)}{\dot r^Q(b_1)}c_0+\frac{\dot r^Q(b_2)-\dot r^Q(a_2)}{\dot r^Q(b_1)}$;

(iii) For
$\dot r^Q(b_1)<0$, $c_0$ and $c_1$ satisfy
$c_1>\frac{\dot r^Q(a_1)}{\dot r^Q(b_1)}c_0+\frac{\dot r^Q(b_2)-\dot r^Q(a_2)}{\dot r^Q(b_1)}$.\\
The common characteristic of the conditions (i)-(iv) is that the valid value ranges of $c_0$ and $c_1$ should control (or smooth) the difference between $|c_1\dot  r^Q(b_1))|$ and $|c_0\dot  r^Q(a_1)|$. From all the observations above, we see that the smoothness condition S0 holds in many cases of value ranges of $\dot  r^Q(a_1)$, $\dot  r^Q(b_1)$, $\dot  r^Q(a_2)$, $\dot  r^Q(b_2)$, $c_0$ and $c_1$.

For general case where $\ddot \xi_o(x)> 0$ (or $\ddot \xi_o(x)< 0$) and $\ddot r^Q(x)>0$ (or $\ddot r^Q(x)<0$) for $x$  in some subintervals of $[0,1]$, the conclusions are similar, but the interpretations are complex.

In the following, we consider the smoothness of basis function representation in (2.11).
Let $\phi_j(\cdot)$ be orthogonal basis functions, satisfying
$$\int_0^1 \phi_j(z)\phi_k(z)dz=0 \mbox{ for }j\neq k, \mbox{ and }\int_0^1 \phi^2_j(z)dz=1.$$  For convenience, we first suppose that
$z=r^P(x)$ is a monotone function in $[0,1]$. The above can be equivalently rewritten as
$$\int_{r^P(0)}^{r^P(1)} \phi_j(r^P(x))\phi_k(r^P(x))dr^P(x)=0 \mbox{ for }j\neq k, \mbox{ and }\int_{r^P(0)}^{r^P(1)} \phi^2_j(r^P(x))dr^P(x)=1.$$
Regarding $\phi_j(r^P(x))$ as orthogonal basis functions, it is reasonable to suppose that $r^Q(x)$ can be expressed as
$$r^Q(x)=\sum_{k=1}^\infty\alpha_k\phi_k(r^P(x)).$$ Similar to the case of linear framework, it holds that $$r^Q(x)\approx\sum_{k=1}^m\xi_{k}(x)\phi_k(r^P(x)).$$
By comparing the two representations above, it follows that
$$\xi_{k}(x)\approx \alpha_k \mbox{ for } k=1,\cdots,m.$$ The above ensures that $\xi_{k}(x),k=1,\cdots,m,$ are constant functions, approximately, implying that $\xi_k(x),k=1,\cdots,m,$ are smoother than $r^Q(x)$.

For the case where $z=r^P(x)$ is a monotone function in some subintervals of $[0,1]$, the conclusion is similar.

Finally, we briefly check the smoothness of  general adjustment function $f(r^P(x),x)$ with $f(0,x)=0$ defined in (2.3). Note that $f(r^P(x),x)$ can be expressed as $f(g(r^Q(x)),x)$ for a smooth function $g(\cdot)$ by the assumption given above, and theoretically, the ideal function $f$ should minimizes the $L_2$-criterion:
$\int_0^1 K_{h^Q}(t-x)\{r^Q(t)-f(g(r^Q(t)),t)\}^2dt$ at each fixed target point $x$. Thus $f(g(r^Q(x)),x)\approx r^Q(x)$ for all $x$.
On the other hand,
at each $x$, $f(g(r^Q(x)),x)$ can be thought of as a function of $g(r^Q(x))$ with $f(0,x)=0$, and then it can be assumed that $f(g(r^Q(x)),x)$ has the following representation:
$$f(g(r^Q(x)),x)=\sum_{j=1}^\infty\vartheta_j(x)\phi_j(g(r^Q(x))),$$
where $\phi_j(x)$ are orthogonal basis functions. Due to $f(g(r^Q(x)),x)\approx r^Q(x)$, each coefficient satisfies $\vartheta_j(x)\approx\int_{g(r^Q(0))}^{g(r^Q(1))} r^Q(x)\phi_j(g(r^Q(x)))dg(r^Q(x))$, a function of $g(r^Q(x))$.
Then $f(g(r^Q(x)),x)$ can be rewritten as $f(g(r^Q(x)),x)=m(g(r^Q(x)))$ for a smooth function $m(u)$. Note that $f(g(r^Q(x)),x)\approx r^Q(x)$ implies that $m(g(r^Q(x)))\approx r^Q(x)$, i.e., $m\circ g(u)\approx u$, an approximate identity function. Therefore, the ideal choice of $f(g(r^Q(x)),x)$ is an identity function, and then is smoother than $r^Q(x)$.

\subsection{Regularity conditions and additional theoretical conclusions}

We first introduce the conditions of kernel function and bandwidth under the nonparametric models:
\begin{itemize}\item[S1.] The kernel function $K(t)$ has the second-order continuous and bounded derivative, and is symmetric with
respect to $t=0$, and satisfies $\int K(t)dt=1$, $\int t^2
K(t)dt<\infty$ and $\int  K^2(t)dt<\infty$.
\item[S2.]
The sample sizes and bandwidths satisfy $h^P,h^Q\rightarrow 0$,
$n^Ph^P\rightarrow \infty$ and $n^Qh^Q\rightarrow \infty$.
\end{itemize}
It is obvious that S1 is a common kernel function condition, and in condition S2, the bandwidth conditions are very common in kernel estimation (see, e.g., \citealp{hart2013nonparametric}).

The following theorem and corollaries are employed as the theoretical support for the method in Subsection 2.3.

\noindent{\bf Theorem S.1.} {\it Under the random design with  conditions C1, in addition to the above regularity conditions S1 and S2, suppose the following condition holds:
\begin{itemize}
\item[S3.] The density functions $\varphi^P(x)$ and $\varphi^Q(x)$ of $X^P$ and $X^Q$ have the second-order continuous and bounded derivatives, and $\varphi^P(x)>0$ and $\varphi^Q(x)>0$ for $x\in[L,U]$.
   \end{itemize}
If the sample sizes and bandwidths satisfy
$h^P=o((n^Q)^{-1/5})$ and $h^Q=o((n^Q)^{-1/5})$, and $n^Ph^P/(n^Qh^Q)\rightarrow \tau> 0$, then, the sw-TLE (2.13) satisfies

(i) for $0<\tau<\infty$ and $x\in(L,U)$,
\begin{equation*}\begin{split}& \sqrt{n^Ph^P+n^Qh^Q}
\left(\widehat {r}_b^Q(x)-r^Q(x)\right)\\&\stackrel{d}\rightarrow
N\left(0,\left(1+\frac{1}{\tau}\right)\left(\frac{2r^Q(x)
}{r^P(x)}\right)^2
\frac{\sigma^2_P}
{\varphi^P(x)}\int_{L}^{U}K^2(t)dt+\frac{\sigma^2_Q(1+\tau)}
{\varphi^Q(x)}\int_{L}^{U}K^2(t)dt
\right);\end{split}\end{equation*}

(ii) for $\tau=\infty$ and $x\in(L,U)$,
\begin{equation*}\begin{split} \sqrt{n^Qh^Q}
\left(\widehat {r}_b^Q(x)-r^Q(x)\right)\stackrel{d}\rightarrow
N\left(0,\frac{\sigma^2_Q}
{\varphi^Q(x)}\int_{L}^{U}K^2(t)dt
\right);\end{split}\end{equation*}

(iii) for $\tau=0$ and $x\in(L,U)$,
\begin{equation*}\begin{split} \sqrt{n^Ph^P}
\left(\widehat {r}_b^Q(x)-r^Q(x)\right) \stackrel{d}\rightarrow
N\left(0, \left(\frac{2r^Q(x)
}{r^P(x)}\right)^2
\frac{\sigma^2_P}
{\varphi^P(x)}\int_{L}^{U}K^2(t)dt\right).\end{split}\end{equation*}

}

\noindent{\bf Corollary S.2.} {\it Under the conditions of Theorem S.1, for the case of $0<\tau<\infty$, the relative efficiency of  the sw-TLE (2.13) can be expressed asymptotically as
$$
\mathbb{RE}[\widehat {r}_b^Q(x),\widehat r^Q(x)]=\frac{\tau\sigma_Q^2\int_L^U
\frac{1}{\varphi^Q(x)}dx}{\sigma_P^2\int_L^U\left(\frac{2r^Q(x)}{r^P(x)}
\right)^2\frac{1}{\varphi^P(x)}dx+\tau\sigma_Q^2\int_L^U
\frac{1}{\varphi^Q(x)}dx}.$$
Consequently, we have $$\mathbb{RE}[\widehat {r}_b^Q(x),\widehat r^Q(x)]> 1$$ if and only if the following condition holds:
\begin{itemize}
\item[S4] $\rho>1$ and $\tau>\frac{\sigma_P^2\int_L^U\left(\frac{2r^Q(x)}{r^P(x)}
\right)^2\frac{1}{\varphi^P(x)}dx}{\sigma_Q^2\left (\rho-1\right)\int_L^U\frac{1}{\varphi^Q(x)}dx}.$
\end{itemize}}

\noindent{\bf Corollary S.3.} {\it Under the conditions of Theorem S.1, for the case of $0<\tau<\infty$, suppose that $n^P/n\rightarrow \tau_P\neq 0$ and $(n^Ph^P+n^Qh^Q)/(nh)\rightarrow\phi$, then,  the sw-TLE (2.13) satisfies
\begin{equation*}\begin{split}&\mathbb{RE}[\widehat {r}_b^Q(x),\widehat r(x)]\rightarrow\infty\mbox{ if } r^P(x)\neq r^Q(x) \mbox{ with } x\in (c_1,c_2)\subset [L,U] \mbox{ for some } c_2>c_1;\\&\mathbb{RE}[\widehat {r}_b^Q(x),\widehat r(x)]=\frac{\rho\phi}
{2+\tau+\frac{1}{\tau}} \mbox{ if P = Q},\label{(efficiency-full-1)}\end{split}\end{equation*}where $\varphi(x)$ is the density function of data set $D_X$.}

\noindent{\bf Theorem S.4.} {\it Under the random design model with the conditions C1', S1, S2 and S3, if the sample sizes and bandwidths satisfy the conditions given in Theorem S.1, then the sw-TLE (2.14) has the following asymptotic normality:

(i) for $0<\tau<\infty$ and $x\in(L,U)$,
\begin{equation*}\begin{split}& \sqrt{n^Ph^P+n^Qh^Q}
\left(\widehat {r}_{bk}^Q(x)-r^Q(x)\right)\\&\stackrel{d}\rightarrow
N\left(0,\left(1+\frac{1}{\tau}\right)\left(2\bm{\xi}^0_k(x))^T
\dot{\bm\phi}_k^P(x)\right)^2
\frac{\sigma^2_P}
{\varphi^P(x)}\int_{L}^{U}K^2(t)dt+\frac{\sigma^2_Q(1+\tau)}
{\varphi^Q(x)}\int_{L}^{U}K^2(t)dt
\right);\end{split}\end{equation*}

(ii) for $\tau=\infty$ and $x\in(L,U)$,
\begin{equation*}\begin{split} \sqrt{n^Qh^Q}
\left(\widehat {r}_{bk}^Q(x)-r^Q(x)\right) \stackrel{d}\rightarrow
N\left(0,\frac{\sigma^2_Q}
{\varphi^Q(x)}\int_{L}^{U}K^2(t)dt
\right);\end{split}\end{equation*}

(iii) for $\tau=0$ and $x\in(L,U)$,
\begin{equation*}\begin{split} \sqrt{n^Ph^P}
\left(\widehat {r}_{bk}^Q(x)-r^Q(x)\right) \stackrel{d}\rightarrow
N\left(0,\left(2\bm{\xi}^0_k(x))^T
\dot{\bm\phi}_k^P(x)\right)^2
\frac{\sigma^2_P}
{\varphi^P(x)}\int_{L}^{U}K^2(t)dt\right).\end{split}\end{equation*}

}

Next, we introduce the regularity conditions for the semiparametric models:

\begin{itemize}
\item[S5.]
The sample size and bandwidth satisfy $h^Q\rightarrow 0$,
and $n^Qh^Q\rightarrow \infty$.
\item[S6.] The regression function
$r^Q(x)$ has the second-order continuous and bounded derivative.
\item[S7.] $\Theta$ is a convex subset of $R^d$, the true value of $\theta=(\theta_1,\cdots,\theta_d)^T$ is an inner point of $\Theta$.
    \item[S8.] The first and second derivatives $\dot r^P(x,\theta)$ and $\ddot r^P(x,\theta)$ of $r^P(x,\theta)$ with respective to $\theta$ exist and are continuous for all  $\theta\in\Theta^0$, where $\Theta^0\subset \Theta$ is an open neighborhood of the true value $\theta^0$ of $\theta$.
\item[S9.] $\frac{1}{n^P}\sum_{i=1}^{n^P}(\dot r^P(X^P_i,\theta))^T\dot r^P(X^P_i,\theta)$ converges in probability to $\mathbb{V}_r^P(\theta)$ for all $\theta\in\Theta^0$, where  $\mathbb{V}_r^P(\theta)=\mathbb{E}[\dot r^P(X^P,\theta)(\dot r^P(X^P,\theta))^T]$ is a nonsingular matrix, and furthermore,  $\frac{1}{n^P}\sum_{i=1}^{n^P}(\partial^2 r^P(X^P_i,\theta) /\partial \theta_j\partial \theta_k)^2$ converges in probability for $j,k=1,\cdots,d$, and all $\theta\in\Theta^0$.
\end{itemize}
These conditions are very common under parametric and nonparametric regression models (see, e.g., \citealp{seber2003nonlinear,hart2013nonparametric}).

\subsection{Multi-source model}

In the following, we list the details about the estimation method and theoretical property under multi-source parametric models:
\begin{equation*}\mbox{(P$_j$-model)} \ \ \ Y_{ij}^P=r_j^P(X_{ij}^P,\theta_j)+\varepsilon_{ij}^P,i=1,
\cdots,n_j^P,j=1,\cdots,m,\end{equation*} where the errors satisfy $\mathbb{E}[\varepsilon_{ij}^P|X_{ij}^P]=0$ and $\mathbb{V}[\varepsilon_{ij}^P|X_{ij}^P]=\sigma^2_{Pj}$. Let $\widehat\theta_j$ be the least squares estimator of $\theta_j$ from P$_j$-model. For each source model, we adjust the estimator $r_j^P(x,\widehat\theta_j)$ to the following form:
$$r_j^P(x,\widehat\theta_j)\widehat\alpha_j(x),$$ where the estimator of the adjustment factor is defined by \begin{equation*}\begin{split}\widehat\alpha_j(x)=
\frac{\sum_{i=1}^{n_j^Q}K_{h^Q}(X_{ij}^Q-x)Y_i^Qr_j^P
(X_{ij}^Q,\widehat\theta_j)}{\sum_{i=1}^{n_j^Q}K_{h^Q}(X_{ij}^Q-x)
(r_j^P(X_{ij}^Q,\widehat\theta_j))^2}.\end{split}\end{equation*}
By combining the above estimators, we attain the sw-TLE of the target function $r^Q(x)$ as
\begin{equation}
\label{equ_muti}
\widehat {r}_c^Q(x)=\sum_{j=1}^mw_jr_j^P(x,\widehat\theta_j)\widehat\alpha_j(x), 
\end{equation}
where weights satisfy $w_j\geq 0$ and $\sum_{j=1}^{m}w_j=1$. Usually, the chosen weights satisfy $w_j\propto n_j^P$ and $w^{-1}_j\propto \mathbb{V}[r_j^P(x,\widehat\theta_j)\widehat\alpha_j(x)]$.

We also can establish its theoretical property.
For example, similar to Theorem 3.1, under condition $n_j^P/(n^Qh^Q)\rightarrow \tau_j$ for some constants $0<\tau_j<\infty$, if
$h^Q=o((n^Q)^{-1/5})$, then, the sw-TLE $\widehat {r}_c^Q(x)$ satisfies
\begin{equation*}\begin{split}&\left(\sqrt{\sum_{j=1}^mn_j^P}+\sqrt{n^Qh^Q}\right)
\left(\widehat {r}_c^Q(x)-r^Q(x)\right)\\&\stackrel{d}\rightarrow
N\left(0,\sum_{j=1}^mw_j^2\left(\sigma^2_{Pj}\left(1+\frac{1}{\tau_j}\right)
\phi_j(\theta^0_j)\left(\frac{2r^Q(x)
}{r^P_j(x,\theta^0_j)}\right)^2+\frac{\sigma^2_Q(\tau_j+1)}
{\varphi^Q(x)}\int_{L}^{U}K^2(x)dx\right)
\right)\end{split}\end{equation*} for all $x\in(L,U)$,
where
$\phi_j(\theta^0_j)=(\mathbb{E}[\dot r^P_j(X^P,\theta^0_j)])^T
(\mathbb{V}_{r_j}^P(\theta^0_j))^{-1}\mathbb{E}[(\dot r^P_j(X^P,\theta^0_j))^P]
$.

\subsection{Proofs}

In the procedures of proving the theorems, we need the Bahadur expresentations
of the estimators from the source model P. For example, as shown by the existing literature (see, e.g., \citealp{bhattacharya1990kernel}; \citealp{chaudhuri1991nonparametric}; \citealp{chaudhuri1991nonparametric}; \citealp{hong2003bahadur}), the Bahadur representation of N-W estimator $\widehat r^P(x)$ can be expressed as
$$\label{(Bahadur)}\widehat r^P(x)=r^P(x)+\beta_{n^P}(x)+\epsilon_{n^P},$$
where $$\beta_{n^P}(x)=\frac{1}{vn^P}\sum_{i=1}^{n^P}K_{h^P}(X_i^P-x)
(Y_i^P-r^P(x))$$
with $v=\int K(t)\varphi^P(x+h^Pt)dt$, and $\epsilon_{n^P}=O_p((n^P)^{-3/4})$ .
Furthermore, in order to prove Theorem 2.1, we first introduce the following lemmas.

\noindent{\bf Lemma S.1.} {\it Under the conditions of Theorem 2.1, we have
$$\widehat\xi(x)=\xi_{h^Q}(x)+O_p\left(\frac{1}{n^Qh^Q}\right).$$
}

\noindent{\it Proof.} We only need to prove
\begin{equation*}\begin{split}&\mathbb{E}\left[\widehat\xi(x)|D^P\right]
=\xi_{h^Q}(x)+O_p\left(\frac{1}{n^Q}\right), \\&  \mathbb{V}\left[\widehat\xi(x)|D^P\right]=\frac{\sigma_Q^2
\int_{0}^{1}K^2(t)
dt}{n^Qh^Q\varphi^Q(x)(r^P(x))^2}+
o_p\left(\frac{1}{n^Qh^Q}\right).\end{split}\end{equation*}
Denote $ I=\sum_{i=1}^{n^Q}Y_i^Q\int_{s_{i-1}^Q}^{s_i^Q}K_{h^Q}(t-x)\widehat {r}^P(t)dt,$ the
numerator of $\widehat\xi(x)$.
By the property of kernel function and the quasi-uniformity, we have
\begin{equation*}\begin{split}\mathbb{E}\left[I|D^P\right]&=\sum_{i=1}^{n^Q}r^Q(X_i^Q)\int_{s_{i-1}^Q}^{s_i^Q}K_{h^Q}(t-x)\widehat {r}^P(t)dt\\&=\sum_{i=1}^{n^Q}\int_{s_{i-1}^Q}^{s_i^Q}K_{h^Q}(t-x)r^Q(t)\widehat {r}^P(t)dt+O(1/n^Q)\\&=\int_{0}^{1}K_{h^Q}(t-x)r^Q(t)\widehat {r}^P(t)dt+O_p(1/n^Q).
\end{split}\end{equation*} It follows from the result above and the definitions of $\xi_{h^Q}(x)$ and $\widehat\xi(x)$ that
$$\mathbb{E}\left[\widehat\xi(x)|D^P\right]=\xi_{h^Q}(x)+O_p(1/n^Q).$$
On the other hand, the conditional variance  \begin{equation*}\begin{split}\mathbb{V}[I|D^P]&=
\sigma_Q^2\sum_{i=1}^{n^Q}
\left(\int_{s_{i-1}^Q}^{s_i^Q}K_{h^Q}(t-x)
\widehat {r}^P(t)dt\right)^2\\&=\sigma_Q^2\sum_{i=1}^{n^Q}
(s_i^Q -s_{i-1}^Q)^2\left(K_{h^Q}(x_i^*-x)
\widehat {r}^P(x_i^*)\right)^2,\end{split}\end{equation*} where $x_i^*\in[s_i^Q,s_{i-1}^Q]$. By the quasi-uniform condition $s_i^Q -s_{i-1}^Q=1/(\varphi^Q(X^Q_i)n^Q)+o(1/n^Q)$, we have
\begin{equation*}\begin{split}\mathbb{V}[I|D^P]&=\sigma_Q^2\sum_{i=1}^{n^Q}
\frac{1}{\varphi^Q(X^Q_i)n^Q}(s_i^Q -s_{i-1}^Q)\left(K_{h^Q}(x_i^*-x)
\widehat {r}^P(x_i^*)\right)^2+o_p(\mathbb{V}[I|D^P])\\&=\sigma_Q^2\sum_{i=1}^{n^Q}
\int_{s_{i-1}^Q}^{s_i^Q}\frac{1}{\varphi^Q(t)n^Q}\left(K_{h^Q}(t-x)
\widehat {r}^P(t)\right)^2dt +o_p(\mathbb{V}[I|D^P])\\&=\frac{\sigma_Q^2
(\widehat {r}^P(x))^2\int_{0}^{1}K^2(t)
dt}{n^Qh^Q\varphi^Q(x)}
+o_p(\mathbb{V}[I|D^P])\\&=\frac{\sigma_Q^2(r^P(x))^2
\int_{0}^{1}K^2(t)
dt}{n^Qh^Q\varphi^Q(x)} +o_p(\mathbb{V}[I|D^P]).\end{split}
\end{equation*}
Then,
\begin{equation*}\begin{split}\mathbb{V}[\widehat\xi(x)|D^P]&=\frac{\sigma_Q^2(r^P(x))^2
\int_{0}^{1}K^2(t)
dt}{n^Qh^Q\varphi^Q(x)(\int_0^1 K_{h^Q}(t-x)
(\widehat {r}^P(t))^2dt)^2}+
o_p\left(1/(n^Qh^Q)\right)\\&=\frac{\sigma_Q^2
\int_{0}^{1}K^2(t)
dt}{n^Qh^Q\varphi^Q(x)(r^P(x))^2}+
o_p\left(1/(n^Qh^Q)\right).\end{split}\end{equation*}
The proof is completed. $\square$

\noindent{\bf Lemma S.2.} {\it Under the conditions of Theorem 2.1, we have
\begin{equation*}\begin{split}\widehat r^P(x)\widehat\xi(x)&= r^P(x)\xi_{h^Q}(x)+\xi_{h^Q}(x)(\widehat r^P(x)-r^P(x))+r^P(x)(\widehat\xi(x)-\xi_{h^Q}(x)) \\&\ \ \ +o_p\left((h^Q)^2+\frac{1}{\sqrt{n^Qh^Q}}\right).\end{split}\end{equation*}
}
\noindent{\it Proof}. By Lemma S.1 and $\widehat r^P(x)-r^P(x)=O_p\left((h^P)^2+1/\sqrt{n^Ph^P}\right)$, we have
\begin{equation*}\begin{split}\widehat r^P(x)\widehat\xi(x)&= r^P(x)\xi_{h^Q}(x)+\xi_{h^Q}(x)(\widehat r^P(x)-r^P(x))+r^P(x)(\widehat\xi(x)-\xi_{h^Q}(x))\\&\ \ \ +(\widehat\xi(x)-\xi_{h^Q}(x))(\widehat r^P(x)-r^P(x))\\&= r^P(x)\xi_{h^Q}(x)+\xi_{h^Q}(x)(\widehat r^P(x)-r^P(x))+r^P(x)(\widehat\xi(x)-\xi_{h^Q}(x)) \\&\ \ \ +o_p\left((h^Q)^2 +1/\sqrt{n^Qh^Q}\right).\end{split}\end{equation*}
The proof is completed.
$\square$

\noindent{\bf Lemma S.3.} {\it
Under the conditions of Theorem 2.1, we have
$$\xi_{h^Q}(x)=\xi^*_{h^Q}(x)
+\left(\frac{r^Q(x)}{({r}^P(x))^2}
-\frac{2\xi^*_{h^Q}(x)}{{r}^P(x)}\right)\beta_{n^P}(x)
+o_p((h^P)^2+1/\sqrt{n^Ph^P}),$$ where $\xi^*_{h^Q}(x)
=\frac{\int_0^1 K_{h^Q}(t-x)r^Q(t){r}^P(t)dt}{\int_0^1 K_{h^Q}(t-x)
({r}^P(t))^2dt}$ and $\beta_{n^P}(x)$ is the
main part of the Bahadur
representation of $\widehat r^P(x)-r^P(x)$ given above.
}

\noindent{\it Proof}. It follows from the property of kernel estimator and the Bahadur representation of $\widehat r^P(x)$ that
\begin{equation*}\begin{split}\xi_{h^Q}(x)&=\frac{\int_0^1 K_{h^Q}(t-x)r^Q(t)\widehat {r}^P(t)dt}{\int_0^1 K_{h^Q}(t-x)
(\widehat{r}^P(t))^2dt}\\&
=\frac{\int_0^1 K_{h^Q}(t-x)r^Q(t)({r}^P(t)+\beta_{n^P}(t))dt}{\int_0^1 K_{h^Q}(t-x)
({r}^P(t)+\beta_{n^P}(x))^2dt}+o_p((h^P)^2+1/\sqrt{n^Ph^P})\\&
=\frac{\int_0^1 K_{h^Q}(t-x)r^Q(t){r}^P(t)dt}{\int_0^1 K_{h^Q}(t-x)
(({r}^P(t))^2+2r^P(x)\beta_{n^P}(x)+o_p(\beta_{n^P}(x)))dt}\\&
\ \ \ +\frac{\int_0^1 K_{h^Q}(t-x)r^Q(t)\beta_{n^P}(t)dt}{\int_0^1 K_{h^Q}(t-x)
(({r}^P(t))^2+o_p(1))dt}+o_p((h^P)^2+1/\sqrt{n^Ph^P})\\&
=\frac{\int_0^1 K_{h^Q}(t-x)r^Q(t){r}^P(t)dt}{\int_0^1 K_{h^Q}(t-x)
({r}^P(t))^2dt}\left(1-\frac{2\int_0^1 K_{h^Q}(t-x)r^P(t)\beta_{n^P}(t)dt}{\int_0^1 K_{h^Q}(t-x)
({r}^P(t))^2dt}\right)\\&
\ \ \ +\frac{\int_0^1 K_{h^Q}(t-x)r^Q(t)\beta_{n^P}(t)dt}{\int_0^1 K_{h^Q}(t-x)
({r}^P(t))^2dt}+o_p((h^P)^2+1/\sqrt{n^Ph^P})\\&
=\xi^*_{h^Q}(x)-\frac{2\xi^*_{h^Q}(x)}{{r}^P(x)}\beta_{n^P}(x)
+\frac{r^Q(x)}{({r}^P(x))^2}\beta_{n^P}(x)
+o_p((h^P)^2+1/\sqrt{n^Ph^P}).\end{split}\end{equation*}
We then complete the proof. $\square$

\noindent{\bf Lemma S.4.} {\it Let $w_i^a(x)=
\frac{\int_{s_{i-1}^Q}^{s_i^Q}K_{h^Q}(t-x)  {r}^P(t)dt}{\int_0^1 K_{h^Q}(t-x)
( {r}^P(t))^2dt}$. Then, under the conditions of Theorem 2.1, we have
\begin{equation*}\label{(relation)}\begin{split}&  r^P(x)(\widehat\xi(x)-\xi_{h^Q}(x))\\&=\sum_{i=1}^{n^Q}
r^P(x) \left(Y^Q_iw_i^a(x)-\xi^*_{h^Q}(x)\right)+ \frac{r^Q(x)}{{r}^P(x)}
\beta_{n^P}(x) +o_p((h^P)^2+1/\sqrt{n^Ph^P}).
\end{split}\end{equation*}}

\noindent{\it Proof.}
Denote $\widehat w_i^a(x)=
\frac{\int_{s_{i-1}^Q}^{s_i^Q}K_{h^Q}(t-x)\widehat {r}^P(t)dt}{\int_0^1 K_{h^Q}(t-x)
(\widehat {r}^P(t))^2dt}.$
By the Bahadur representation of $\widehat r^P(x)$, we have
\begin{equation*}\begin{split}
 \widehat w_i^a(x)&
=\frac{
\int_{s_{i-1}^Q}^{s_i^Q}K_{h^Q}(t-x) ({r}^P(t)+\beta_{n^P}(t))dt}{\int_{0}^{1}K_{h^Q}(t-x)
(\widehat{r}^P(t))^2dt}+O_p\left(\frac{1}{n^Q}(n^Q)^{-3/4}\right)\\&= \left( w_i^a(x)
+\frac{\int_{s_{i-1}^Q}^{s_i^Q}K_{h^Q}(t-x)\beta_{n^P}(t)dt}
{\int_{0}^{1}K_{h^Q}(t-x)
({r}^P(t))^2dt}\right) \left(1+o_p(1)\right)+O_p\left(\frac{1}{n^Q}(n^Q)^{-3/4}\right)
\\&= w_i^a(x)+O_p\left(\frac{1}{n^Q}((h^Q)^2+1/\sqrt{n^Qh^Q})\right).\end{split}\end{equation*}
Then, by the result above and Lemma S.3, we have
\begin{equation*}\label{(relation)}\begin{split}&  r^P(x)(\widehat\xi(x)-\xi_{h^Q}(x)) \\&=\sum_{i=1}^{n^Q}
r^P(x)Y^Q_i\widehat w_i^a(x)- r^P(x)\xi_{h^Q}(x) \\&=\sum_{i=1}^{n^Q}
r^P(x) Y^Q_iw_i^a(x) -r^P(x)\xi^*_{h^Q}(x) - r^P(x)\left(\frac{r^Q(x)}{({r}^P(x))^2}
-\frac{2\xi^*_{h^Q}(x)}{{r}^P(x)}\right)\beta_{n^P}(x)\\& \ \ \ +o_p((h^P)^2+1/\sqrt{n^Ph^P})\\&=\sum_{i=1}^{n^Q}
r^P(x) Y^Q_iw_i^a(x) -r^P(x)\xi^*_{h^Q}(x) + \frac{r^Q(x)}{{r}^P(x)}
\beta_{n^P}(x) +o_p((h^P)^2+1/\sqrt{n^Ph^P}).
\end{split}\end{equation*}
The proof is completed. $\square$

\

\noindent{\it Proof of Theorem 2.1.} We only prove the case of $ 0<\tau<\infty$. For the other cases, the proofs are similar. Because of the under-smoothing condition
of $h^P=o((n^O)^{-1/5})$ and $h^Q=o((n^Q)^{-1/5})$, all the asymptotic biases of order
$O_p((h^P)^2)$ and $O_p((h^Q)^2)$ can be
ignored for establishing the asymptotic normality. By Lemmas S.1-Lemma S.4, we have
\begin{equation*}\begin{split}&
\widehat {r}^P(x)\widehat\xi(x)-r^Q(x)\\&=
\xi_{h^Q}(x)(\widehat r^P(x)-r^P(x))+r^P(x)(\widehat\xi(x)-\xi_{h^Q}(x))+r^P(x)\xi_{h^Q}(x)-r^Q(x) \\&\ \ \ +o_p\left((h^Q)^2+1/\sqrt{n^Qh^Q}\right)\\&=
\xi_{h^Q}(x)(\widehat r^P(x)-r^P(x))+r^P(x)(\widehat\xi(x)-\xi_{h^Q}(x))+r^P(x)\xi^*_{h^Q}(x)-r^Q(x) \\&\ \ \ +o_p\left((h^Q)^2+1/\sqrt{n^Qh^Q}\right)\\&=
\xi_{h^Q}(x)(\widehat r^P(x)-r^P(x))+r^P(x)(\widehat\xi(x)-\xi_{h^Q}(x)) +O((h^Q)^2)+o_p\left((h^Q)^2+1/\sqrt{n^Qh^Q}\right).\end{split}\end{equation*}
Note that $\xi_{h^Q}(x)\rightarrow \frac{r^Q(x)}{r^P(x)}$ in probability.
The above results and Lemma S.4 lead to
\begin{equation*}\begin{split}&\sqrt{n^Ph^P+n^Qh^Q}
\left(\widehat {r}^P(x)\widehat\xi(x)-r^Q(x)\right)\\&=\sqrt{n^Ph^P+n^Qh^Q}
\left(\xi_{h^Q}(x)(\widehat r^P(x)-r^P(x))\right) +\sqrt{n^Ph^P+n^Qh^Q}r^P(x)(\widehat\xi(x)-\xi_{h^Q}(x)) +o_p\left(1\right)\\&=\sqrt{1+\frac{1}{\tau}}\sqrt{n^Ph^P}
\left(\xi_{h^Q}(x)(\widehat r^P(x)-r^P(x))+ \frac{r^Q(x)}{{r}^P(x)}
\beta_{n^P}(x)\right) \\& \ \ \ +\sqrt{1+\tau}\sqrt{n^Qh^Q}r^P(x)\sum_{i=1}^{n^Q}
 \left(Y^Q_iw_i^a(x)
-\xi^*_{h^Q}(x)\right) +o_p\left(1\right)\\&=\sqrt{1+\frac{1}{\tau}}\sqrt{n^Ph^P}
\left(\frac{2r^Q(x)}{r^P(x)}(\widehat r^P(x)-r^P(x))\right) \\& \ \ \ +\sqrt{1+\tau}\sqrt{n^Qh^Q} r^P(x)\sum_{i=1}^{n^Q}
 \left(Y^Q_iw_i^a(x)
-\xi^*_{h^Q}(x)\right) + o_p\left(1\right) \\&\stackrel{\Delta}=\Gamma_1^P
+\Gamma_2^Q+o_p\left(1\right).\end{split}\end{equation*} It can be seen that $\Gamma_1^P$ and $\Gamma_2^Q$
are weighted sum of $Y^P_i,i=1,\cdots,n^P$ and $Y^Q_i,i=1,\cdots,n^Q$, respectively. Thus they are normally distributed
asymptotically. Moreover,
$\Gamma_1^P$ and $\Gamma_2^Q$ are independent of each other due
to the dependence between $D^P$ and $D^Q$.
We then only need to calculate their asymptotic
expectations and variances.

It is known by the property of kernel estimation that the main part of the expectation of
$\widehat r^P(x)-r^P(x)$
is $\frac{(h^P)^2\ddot r^P(x)}{2}\int_0^1t^2K(t)dt$. Then
$$\mathbb{E}[\Gamma_1^P]= \sqrt{1+\frac{1}{\tau}}\sqrt{n^Ph^P}\frac{2r^Q(x)}{r^P(x)}
\frac{(h^P)^2\ddot r^P(x)}{2}\int_0^1t^2K(t)dt+o(1),$$ which can be
ignored because the nonparametric estimator is under-smooth. Furthermore,
\begin{equation*}\begin{split}\left(1+\frac{1}{\tau}\right)n^Ph^P\mathbb{V}[\Gamma_1^P]
=\left(1+\frac{1}{\tau}\right)\left(\frac{2r^Q(x)
}{r^P(x)}\right)^2\frac{\sigma^2_P}
{\varphi^P(x)}\int_{0}^{1}K^2(t)dt+o(1).
\end{split}\end{equation*}

Similarly, $\mathbb{E}[\Gamma_2^Q]$ can be
ignored because of the under-smoothing, and
$$(1+\tau)n^Qh^Q\mathbb{V}[\Gamma_2^Q]=(1+\tau)\frac{\sigma^2_Q}
{\varphi^Q(x)}\int_{0}^{1}K^2(t)dt+o(1).$$
The proof is completed.
$\square$

\

\noindent{\it Proof of Corollary 2.2.} It is a direct result of Theorem 2.1. $\square$

\noindent{\it Proof of Corollary 2.3.} It is a direct result of Theorem 2.1. $\square$

\noindent{\it Proof of Corollary 2.4.} Under fixed design models, the full data estimator is $\widehat r(x)=\sum_{i=1}^nY_i\int _{s_{i-1}}^{s_i}K_h(t-x)dt$ with $(X_i,Y_i)\in D^P\cup D^Q$.
Then, by the property of kernel function and the quasi-uniform condition, we have
\begin{equation*}\begin{split}&\mathbb{E}[\widehat r(x)]
=\sum_{(X_i,Y_i)\in D^P}r^P(X_i)\int _{s_{i-1}}^{s_i}K_h(t-x)dt+\sum_{(X_i,Y_i)\in D^Q}r^Q(X_i)\int _{s_{i-1}}^{s_i}K_h(t-x)dt\\&
=\sum_{(X_i,Y_i)\in D^P}r^P(X_i)({s_i}-s_{i-1})K_h(X^{P*}_i-x)+\sum_{(X_i,Y_i)\in D^Q}r^Q(X_i)({s_i}-s_{i-1})K_h(X^{Q*}_i-x)\\&
=\sum_{(X_i,Y_i)\in D^P}r^P(X_i)\frac{1}{n\varphi(X_i)}K_h(X^{P*}_i-x)+\sum_{(X_i,Y_i)\in D^Q}r^Q(X_i)\frac{1}{n\varphi(X_i)}K_h(X^{Q*}_i-x)+o(1)\\&\rightarrow \tau_P r^P(x)+(1-\tau_P) r^Q(x),\end{split}\end{equation*} where $X^{P*}\in[s_{i-1},s_i]$ with the corresponding sample points $(X_i,Y_i)\in D^P$, and $X^{Q*}\in[s_{i-1},s_i]$ with the corresponding sample points $(X_i,Y_i)\in D^Q$.
It can be seen from the result above that the full data estimator $\widehat r(x)$ has a non-negligible asymptotic bias if $\tau_P\neq 0$ and $r^P(x)\neq r^Q(x)$ for $x\in (c_1,c_2)$. Specifically, the asymptotic bias is $\tau_P r^P(x)+(1-\tau_P) r^Q(x)$, implying $\int_0^1\mathbb{B}^2[\widehat r(x)]dx\nrightarrow 0$ if $\tau_P\neq 0$ and $r^P(x)\neq r^Q(x)$ for $x\in (c_1,c_2)$. Then $\mathbb{RE}[\widehat {r}_a^Q,\widehat r(x)]\rightarrow\infty$.

Contrarily, if $r^P(x)= r^Q(x)$ for all $x$ (i.e., model P and model Q are equal to each other) together with the condition of  under-smoothing,
the asymptotic bias is ignorable.
On the other hand, it is known that
$\mathbb{V}[\widehat r(x)]=\frac{\sigma^2\int_0^1 K^2(t)dt}{nh \varphi(x)}$ asymptotically. This result and Theorem 2.1  together  imply the second result in Corollary 2.4.
 $\square$

\noindent{\it Proof of Theorem 2.5.} We only prove the case of $ 0<\tau<\infty$. For the other cases, the proofs are similar. Note that $\bm{\xi}_{k,h^Q}(x)$ is the minimizer of the local $L_2$-criterion. It satisfies
$$\int_0^1 K_{h^Q}(t-x)\{r^Q(t)-(\widehat{\bm\phi}_k^P(t))^T\bm{\xi}_{k,h^Q}(x)\}
\widehat{\bm\phi}_k^P(t)dt=0.$$
By the above equation and the method of proving Theorem 2.1, we have
\begin{equation*}\begin{split}
r^Q(x){\mathbf{r}}_k^P(x)=({\bm\phi}_k^P(x))^T\bm{\xi}_{k,h^Q}(x) {\bm\phi}_k^P(x)+o_p\left((h^Q)^2+1/\sqrt{n^Qh^Q}\right),\end{split}\end{equation*}
and then
\begin{equation*}\begin{split}
r^Q(x)\|{\bm\phi}_k^P(x)\|^2=({\bm\phi}_k^P(x))^T\bm{\xi}_{k,h^Q}(x)\| {\bm\phi}_k^P(x)\|^2+o_p\left((h^Q)^2+1/\sqrt{n^Qh^Q}\right),\end{split}\end{equation*}
The above result and $\|{\bm\phi}_k^P(x)\|^2\neq 0$ lead to \begin{equation*}\begin{split}r^Q(x)=( {\bm\phi}_k^P(x))^T\bm{\xi}_{k,h^Q}(x)
+o_p\left((h^Q)^2+1/\sqrt{n^Qh^Q}\right).\end{split}\end{equation*}
Thus, similar to the proof of Theorem 2.1, we have
\begin{equation*}
\begin{split}&
(\widehat{\bm{\xi}}_k(x))^T\widehat{\bm\phi}_k^P(x)-r^Q(x)\\&=
(\bm{\xi}_{k,h^Q}(x))^T(\widehat{\bm\phi}_k^P(x)- {\bm\phi}_k^P(x) )+({\bm\phi}_k^P(x))^T(\widehat{\bm{\xi}}_k(x)-\bm{\xi}_{k,h^Q}(x)) \\& \ \ \ +o_p\left((h^Q)^2+1/\sqrt{n^Qh^Q}\right).
\end{split}
\end{equation*}
Note that by the method in the proof of Theorem 2.1, we have
$$ \bm{\xi}_{k,h^Q}(x)=\bm{\xi}^0_k(x)+o_p\left((h^Q)^2+1/\sqrt{n^Qh^Q}\right),$$
where
$\bm{\xi}^0_k(x)=r^Q(x)( {\bm\phi}_k^P(x)({\bm\phi}_k^P(x))^T)^+ {\bm\phi}_k^P(x)$ with $( {\bm\phi}_k^P(x)({\bm\phi}_k^P(x))^T)^+$ being the Moore-Penrose generalized inversion of $ {\bm\phi}_k^P(x)({\bm\phi}_k^P(x))^T$. It can be verified by the definition of Moore-Penrose generalized inversion that $({\bm\phi}_k^P(x)({\bm\phi}_k^P(x))^T)^+=\frac{1}{\|{\bm\phi}_k^P(x)\|^4} {\bm\phi}_k^P(x)({\bm\phi}_k^P(x))^T$. Then, $\bm{\xi}^0_k(x)=\frac{r^Q(x)}{\| {\bm\phi}_k^P(x)\|^2}{\bm\phi}_k^P(x)$.
Consequently,
\begin{equation*}\begin{split}&\sqrt{n^Ph^P+n^Qh^Q}
\left((\widehat{\bm{\xi}}_k(x))^T\widehat{\bm\phi}_k^P(x)-r^Q(x)\right)\\
&=\sqrt{n^Ph^P+n^Qh^Q}(\bm{\xi}_{k,h^Q}(x))^T(\widehat{\bm\phi}_k^P(x)-{\bm\phi}_k^P(x)) +\sqrt{n^Ph^P+n^Qh^Q}({\bm\phi}_k^P(x))^T
(\widehat{\bm{\xi}}_k(x)-\bm{\xi}_{k,h^Q}(x))\\& \ \ \  +o_p\left(1\right)\\&=\sqrt{1+\frac{1}{\tau}}\sqrt{n^Ph^P}
\bm{\xi}^0_k(x)(\widehat{\bm\phi}_k^P(x)-{\bm\phi}_k^P(x)) +\sqrt{1+\tau}\sqrt{n^Qh^Q}({\bm\phi}_k^P(x))^T
(\widehat{\bm{\xi}}_k(x)-\bm{\xi}^0_k(x))\\& \ \ \ +o_p\left(1\right).\end{split}\end{equation*}
Then, by the method of proving Theorem 2.1 and the continuous mapping theorem, we can prove the theorem.
$\square$

In order to prove Theorem S.1, we need the following lemmas. Since the proofs for the lemmas are similar to those of Lemmas S.1-S.4, as an example, only the proof for Lemma S.7 is presented below.

\noindent{\bf Lemma S.5.} {\it Under the conditions of Theorem S.1, we have
$$\widehat\eta(x)=\eta_{h^Q}(x)+O_p((h^P)^2+1/\sqrt{n^Ph^P}).$$}
\noindent{\it Proof}. By the convergence property of kernel estimation,
we can prove the lemma. $\square$

\noindent{\bf Lemma S.6.} {\it Under the conditions of Theorem S.1, we have
\begin{equation*}\begin{split}\widehat r^P(x)\widehat\eta(x)&
= r^P(x)\eta_{h^Q}(x)+\eta_{h^Q}(x)(\widehat r^P(x)-r^P(x))+r^P(x)(\widehat\eta(x)-\eta_{h^Q}(x)) \\&\ \ \ +o_p\left((h^Q)^2+1/\sqrt{n^Qh^Q}\right).\end{split}\end{equation*}
}
\noindent{\it Proof}. By  the same method of proving Lemma S.2, we can prove the lemma. $\square$

\noindent{\bf Lemma S.7.} {\it Denote $\eta^*_{h^Q}(x)
=\frac{\mathbb{E}[K_{h^Q}(X^Q-x)r^Q(X^Q){r}^P(X^Q)]}
{\mathbb{E}[K_{h^Q}(X^Q-x)({r}^P(X^Q))^2]}$, \begin{equation*}\begin{split} c_i^b(x)
=\frac{K_{h^Q}(X_i^Q-x)r^Q(X_i^Q)\beta_{n^P}(X_i^Q)}
{n^Q\mathbb{E}[K_{h^Q}(X^Q-x)({r}^P(X^Q))^2]} \ \mbox{ and } \ d_i^b(x)=
\frac{K_{h^Q}(X_i^Q-x)r^P(X_i^Q)\beta_{n^P}(X_i^Q)}
{n^Q\mathbb{E}[K_{h^Q}(X^Q-x)({r}^P(X^Q))^2]}.\end{split}\end{equation*}
Then, under the conditions of Theorem S.1, we have
$$\eta_{h^Q}(x)=\eta^*_{h^Q}(x)
+\left(\frac{r^Q(x)}{({r}^P(x))^2}
-\frac{2\eta^*_{h^Q}(x)}{{r}^P(x)}\right)\beta_{n^P}(x)
+o_p((h^P)^2+1/\sqrt{n^Ph^P}).$$
}

\noindent{\it Proof}. It follows from the property of kernel estimator and the Bahadur representation of $r^P(x)$ that
\begin{equation*}\begin{split}\eta_{h^Q}(x)&=
\frac{\mathbb{E}[K_{h^Q}(X^Q-x)r^Q(X^Q)\widehat{r}^P(X^Q)]}
{\mathbb{E}[K_{h^Q}(X^Q-x)(\widehat{r}^P(X^Q))^2]} \\&
=\frac{\mathbb{E}[K_{h^Q}(X^Q-x)r^Q(X^Q)({r}^P(X^Q)+\beta_{n^P}(X^Q))]}
{\mathbb{E}[K_{h^Q}(X^Q-x)({r}^P(X^Q)+\beta_{n^P}(X^Q))^2]}
+o_p((h^P)^2+1/\sqrt{n^Ph^P})\\&
=\frac{\mathbb{E}[K_{h^Q}(X^Q-x)r^Q(X^Q){r}^P(X^Q)]}
{\mathbb{E}[K_{h^Q}(X^Q-x)(({r}^P(X^Q))^2+2{r}^P(X^Q)\beta_{n^P}(X^Q)+o_p(\beta_{n^P}(X^Q)]}\\&
\ \ \ +\frac{\mathbb{E}[K_{h^Q}(X^Q-x)r^Q(X^Q)\beta_{n^P}(X^Q)]}
{\mathbb{E}[K_{h^Q}(X^Q-x)(({r}^P(X^Q))^2+o_p(1))]}
+o_p((h^P)^2+1/\sqrt{n^Ph^P})\\&
=\frac{\mathbb{E}[K_{h^Q}(X^Q-x)r^Q(X^Q){r}^P(X^Q)]}
{\mathbb{E}[K_{h^Q}(X^Q-x)({r}^P(X^Q))^2]}
\left(1-\frac{2\mathbb{E}[K_{h^Q}(X^Q-x)r^P(X^Q)\beta_{n^P}(X^Q)]}
{\mathbb{E}[K_{h^Q}(X^Q-x)(({r}^P(X^Q))^2]}\right)\\&
\ \ \ +\frac{\mathbb{E}[K_{h^Q}(X^Q-x)r^Q(X^Q)\beta_{n^P}(X^Q)]}
{\mathbb{E}[K_{h^Q}(X^Q-x)({r}^P(X^Q))^2]}+o_p((h^P)^2+1/\sqrt{n^Ph^P})\\&
=\eta^*_{h^Q}(x)-\frac{2\eta^*_{h^Q}(x)}{{r}^P(x)}\beta_{n^P}(x)
+\frac{r^Q(x)}{({r}^P(x))^2}\beta_{n^P}(x)+o_p((h^P)^2+1/\sqrt{n^Ph^P}).\end{split}\end{equation*}
We then complete the proof. $\square$

\noindent{\bf Lemma S.8.} {\it Let $w_i^b(x)=
\frac{K_{h^Q}(X_i^Q-x)Y_i^Qr^P(X_i^Q)}
{n^Q\mathbb{E}[K_{h^Q}(X^Q-x)
({r}^P(X^Q))^2]}$. Then, under the conditions of Theorem S.1, we have
\begin{equation*}\label{(relation)}\begin{split}&  r^P(x)(\widehat\eta(x)-\eta_{h^Q}(x))\\&=\sum_{i=1}^{n^Q}
r^P(x) \left(Y^Q_iw_i^b(x)-\eta^*_{h^Q}(x)\right)+ \frac{r^Q(x)}{{r}^P(x)}
\beta_{n^P}(x) +o_p((h^P)^2+1/\sqrt{n^Ph^P}).
\end{split}\end{equation*}}

\noindent{\it Proof.} The proof is similar to that of Lemma S.4.

\noindent{\it Proof of Theorem S.1.} By Lemmas S.5-S.8 and the same method of
proving Theorem 2.1, we can prove the theorem.
$\square$

\noindent{\it Proof of Corollary S.2.} It is a direct result of Theorem S.1. $\square$

\noindent{\it Proof of Corollary S.3.}
Under random design models, the full data estimator $\widehat r(x)=\frac{\sum_{i=1}^nY_iK_h(X_i-x)}{\sum_{i=1}^n K_h(X_i-x)}$ with $(X_i,Y_i)\in D^P\cup D^Q$. Write
\begin{equation*}\begin{split}&\sum_P=\sum_{(X^P_i,Y^P_i)\in D^P}Y^P_iK_{h}(X^P_i-x),\sum_Q=\sum_{(X^Q_i,Y^Q_i)\in D^Q}Y^Q_iK_{h}(X^Q_i-x),\\&\sum_P^*=\sum_{(X^P_i,Y^P_i)\in D^P}K_{h}(X^P_i-x),\sum_Q^*=\sum_{(X^Q_i,Y^Q_i)\in D^Q}K_{h}(X^Q_i-x).
\end{split}\end{equation*}
Then the full data estimator can be expressed as  \begin{equation*}\begin{split}\widehat r(x)&=\frac{\sum\limits_P+\sum\limits_Q}{\sum\limits_P^*+\sum\limits_Q^*}
=\frac{\sum\limits_P}{\sum\limits_P^*(1+\sum\limits_Q^*/\sum\limits_P^*)}+
\frac{\sum\limits_Q}{\sum\limits_Q^*(\sum\limits_P^*/\sum\limits_Q^*+1)}.
\end{split}\end{equation*}By the above expression and the property of kernel estimation, we have
$$\widehat r(x)\rightarrow_p\frac{\tau_P\varphi^P(x)}
{\tau_P\varphi^P(x)+(1-\tau_P)\varphi^Q(x)}r^P(x)
+\frac{(1-\tau_P)\varphi^Q(x)}
{\tau_P\varphi^P(x)+(1-\tau_P)\varphi^Q(x)}r^Q(x).$$
It can be seen from the result above that the full data estimator $\widehat r(x)$ has a non-negligible asymptotic bias if $\tau_P\neq 0$ and $r^P(x)\neq r^Q(x)$ for $x\in (c_1,c_2)$. Specifically, the asymptotic bias is
$\frac{\tau_P\varphi^P(x)}
{\tau_P\varphi^P(x)+(1-\tau_P)\varphi^Q(x)}r^P(x)
+(1-\frac{(1-\tau_P)\varphi^Q(x)}
{\tau_P\varphi^P(x)+(1-\tau_P)\varphi^Q(x)})r^Q(x)$, implying $\int_L^U\mathbb{B}^2[\widehat r(x)]dx\nrightarrow 0$ if $\tau_P\neq 0$ and $r^P(x)\neq r^Q(x)$ for $x\in (c_1,c_2)$. In this case, $\mathbb{RE}[\widehat {r}_b^Q,\widehat r(x)]\rightarrow\infty$.

Contrarily, if $r^P(x)= r^Q(x)$ for all $x$ (i.e., model P and model Q are equal to each other),
the asymptotic bias is ignorable due to the condition of  under-smoothing.
On the other hand, it is known that
$\mathbb{V}[\widehat r(x)]=\frac{\sigma^2\int_L^U K^2(t)dt}{nh \varphi(x)}$ asymptotically. This result and Theorem S.1  together  imply the second result in Corollary S.3.
$\square$

\noindent{\it Proof of Theorem S.4.}
By the method of proving Theorem 2.5, we can prove the theorem. $\square$

\noindent{\it Proof of Theorem 3.1.} By the asymptotic property of parameter estimation, we have \begin{equation*}
\begin{split}\sqrt{n^P}
\left(r^P(x,\widehat\theta)-r^P(x,\theta^0)\right)\stackrel{d}\rightarrow
N\left(0,\sigma^2_P(\mathbb{E}[\dot r^P(x,\theta^0)])^T\mathbb{V}^{-1}
(\theta^0)\mathbb{E}[\dot r^P(x,\theta^0)]
\right).\end{split}\end{equation*} This, together with the method in
the proofs of Theorem 2.1 and Theorem S.1, can prove the result of the theorem.
$\square$

\noindent{\it Proof of Corollary 3.2.} It is a direct result of Theorem 3.1. $\square$

\noindent{\it Proof of Theorem 3.3.} Under semiparametric models, we have similar results as in Lemmas S.1-S.5. Note that the convergence rate of parametric estimation is faster than that of nonparametric one. Then, $\sqrt{n^Qh^Q}
\left(r^P(x,\widehat\theta)\widehat\alpha(x)-r^Q(x)\right)=o_p(1)$.
By this result and the same argument as used in the proof of Theorem 2.1, we have
\begin{equation*}\begin{split}&\sqrt{n^Qh^Q}
\left(r^P(x,\widehat\theta)\widehat\alpha(x)-r^Q(x)\right)
\\&=\sqrt{n^Qh^Q}
\left(\alpha^*_{h^Q}(x)(r^P(x,\widehat\theta)-r^P(x,\theta^0))\right) +\sqrt{n^Qh^Q}r^P(x,\theta^0)(\widehat\alpha(x)-\alpha^*_{h^Q}(x)) +o_p\left(1\right)\\&=\sqrt{n^Qh^Q}r^P(x,\theta^0)(\widehat\alpha(x)
-\alpha^*_{h^Q}(x)) +o_p\left(1\right),\end{split}\end{equation*} where $\alpha^*_{h^Q}(x)=
\frac{\mathbb{E}[K_{h^Q}(X^Q-x)Y^Qr^P(X^Q,\theta^0)]}
{\mathbb{E}[K_{h^Q}(X^Q-x)
(r^P(X^Q,\theta^0))^2]}$.
Then, we can prove that theorem.
 $\square$

 \

\subsection{The source-function weighted structure of the James-Stein estimator}  As shown by (2.7), our estimator has a source-function weighted structure. In the following, we check if the James-Stein estimator also has the structure of ``source-function weighting" in the scenario of transfer learning. Here we refer to the perspective of \cite{stigler19901988} to find out this structure.

Consider a simple situation: a collection of independent measurements $X_1, \cdots, X_k$ is available, each measuring a different parameter  $\theta_i$, and each normally distributed $N(\theta_i, 1)$. We then write $X_i=\theta_i+\varepsilon_i$ with $\varepsilon_i\sim N(0,1)$. The ``ordinary" estimator of $\theta_i$ is $\widehat\theta^0_i=X_i$, and the James-Stein estimator is defined by
$$\widehat\theta^{JS}_i=\left(1-\frac{c}{S^2}\right)X_i,$$ where $S^2=\sum_{i=1}^kX_i^2$, and the constant $0<c<2(k-2)$.
Note that ``ordinary"
$\widehat\theta^0_i=X_i$ is inadmissible if $k\geq 3$, because the the James-Stein estimator has uniformly smaller risk for all $\theta_i$, where the risk is defined as
$$\mathbb R(\bm\theta,\widehat{\bm\theta})=\mathbb{E}\left
[\sum_{i=1}^k(\widehat\theta_i-\theta_i)^2\right]$$ with $\bm\theta=(\theta_1,\cdots,\theta_k)^T$ and $\widehat{\bm\theta}=(\widehat\theta_1,\cdots,\widehat\theta_k)^T.$ For a simple proof and explanation, see \cite{stigler19901988}.
This shows that although each $\theta_i$ is unrelated to $(\theta_j,X_j)$ for $j\neq i$, by the information of all the variables $X_1,\cdots,X_k$, instead of single variable $X_i$, the James-Stein estimator is better than the ``ordinary" estimator $\widehat\theta_i^0=X_i$, which only uses the information of $X_i$.

Formally, the James-Stein estimator $\widehat\theta^{JS}_i$ is not of the form of ``weighted sum" of some unrelated variables. However, the James-Stein estimator can be derived by this form, by considering $\theta_i$ as a random variable, and the class of regression estimators of $\mathbb{E}[\theta_i|X_i]$ that are linear in $X_i$ with zero intercept as
$$\widehat\theta_i=bX_i,\ i=1,\cdots,k.$$ For the above linear estimation, by minimizing the lost function
$$L(\bm\theta,\widehat{\bm\theta})=\sum_{i=1}^k(\theta_i-\widehat\theta_i)^2,$$ we get the least squares estimator of $b$ as
\begin{equation}\label{JS} \widehat \beta=\frac{\sum_{i=1}^k\theta_iX_i}{\sum_{i=1}^kX_i^2}\end{equation} if $\theta_i$ can be observed. Thus, the estimator $\widehat \beta$ is a ``weighted sum of $\theta_i$" with weights $X_i$. But each $\theta_i$ is in fact unknown, we need to approximate the ``estimator" $\widehat\beta$. We can estimate $\sum_{i=1}^k\theta_iX_i$ by $\sum_{i=1}^kX_i^2-k$ because $X_i=\theta_i+\varepsilon_i$, and then $\sum_{i=1}^k\theta_iX_i$ and $\sum_{i=1}^kX_i^2-k$ have the same conditional expectation $\sum_{i=1}^k\theta_i^2$, given $\theta_i$. When $\sum_{i=1}^k\theta_iX_i$ is replaced by its estimation $\sum_{i=1}^kX_i^2-k$, we have
$$\widehat\theta_i=\widehat\theta^{JS}_i,\ i=1,\cdots,k.$$ Thus, the James-Stein estimator $\widehat\theta^{JS}_i$ has a hidden structure of weighted sum of $\theta_i$ with weights $X_i$.

In the scenario of transfer learning, the observation $X_i$ can be thought of as a source, and the parameter $\theta_i$ can be regarded as a target.
Thus, the estimator (\ref{JS}) is of the source-function weighted framework. In the case of transfer learning, however, the relationship $X_i=\theta_i+\varepsilon_i$ is not necessarily true. We then need some similarity conditions, for example, $\sum_{i=1}^k\mathbb{E}[X_i^2|\theta_i]-c =\sum_{i=1}^k\theta_i^2$ for some suitable constant $c>0$, to guarantee
$$\widehat\theta_i\approx\widehat\theta^{JS}_i,\ i=1,\cdots,k.$$ Particularly, when
$0<c<2(k-2)$, the following holds:
$$\widehat\theta_i=\widehat\theta^{JS}_i,\ i=1,\cdots,k.$$
Thus, in the scenario of transfer learning, the James-Stein estimator $\widehat\theta^{JS}_i$ and the source-function weighted estimator $\widehat\theta_i$ have similar behavior and structure. This verifies that the James-Stein estimator has a hidden structure of source-function weighting.

\setcounter{equation}{0}
\section{ Further Simulation}
\begin{figure*}[t]
\centering
\begin{tabular}{cc}
\includegraphics[width=0.4\textwidth]{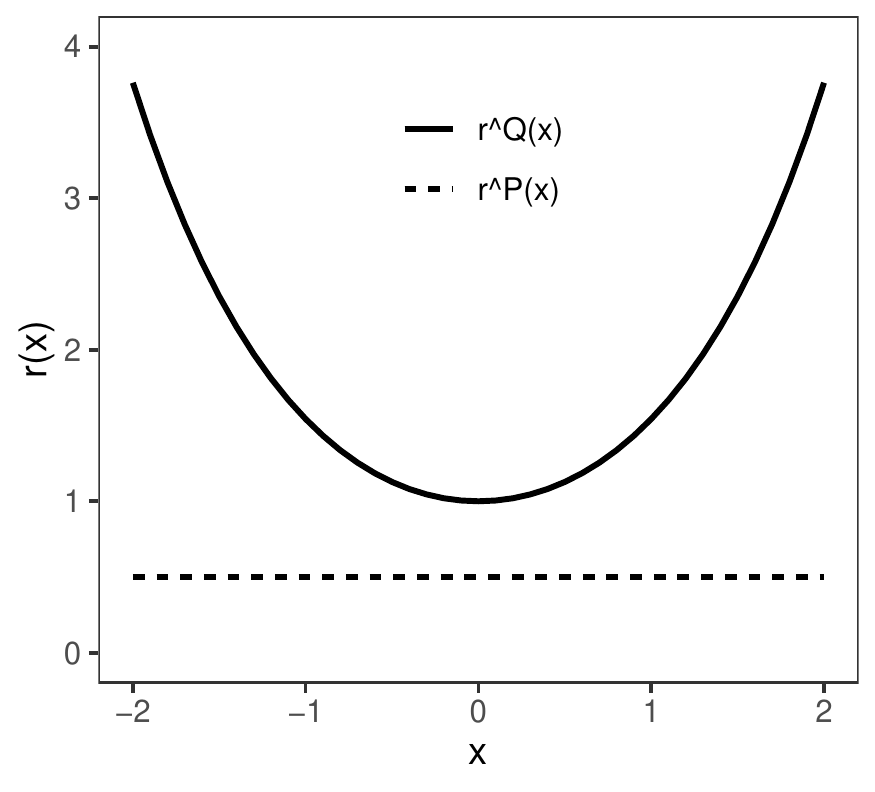}&
\includegraphics[width=0.4\textwidth]{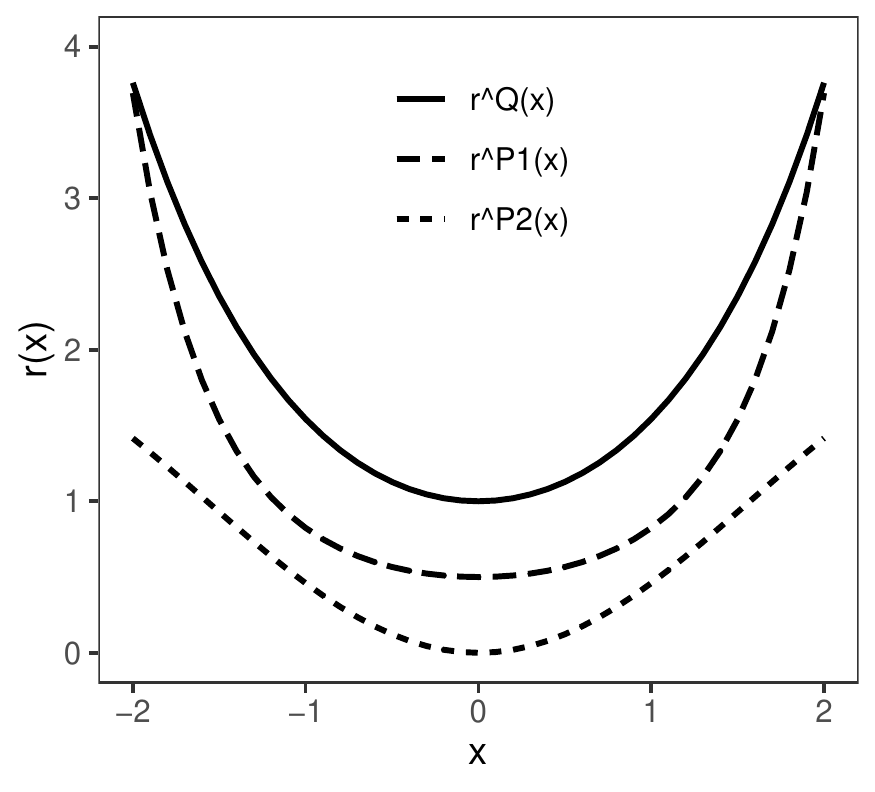}\\
1) unrelated source& 2) multi-source
\end{tabular}
\caption{\small Function graphs: subfigure 1) and 2) are separately for the unrelated source problem and multi-source problem. } \label{fig:examples}
\end{figure*}

In this section, we show more simulation results. The proposed sw-TLE is further analysed for the identical source problem, the unrelated source problem (Fig. \ref{fig:examples}.1) and the multi-source problem  (Fig. \ref{fig:examples}.2).

In the simulation, we consider separately two cases: $n^P=500$ is fixed with varying $n^Q$ and $n^Q=50$ is fixed with varying $n^P$  to show the influence of the data size. Our sw-TLE is compared with the following three  methods:
\begin{itemize}
  \item[1)] The N-W regression. In the identical source problem it is denoted by F-NW as both the source data and target data are employed, while it is denoted by Q-NW in the other two situations in which only the target data is utilized.
  \item[2)] The simple average of the N-W estimator denoted by SA and defined by
  $$
  \widehat{r}^Q_{SA}(x)=\frac{\sqrt{n^Ph^P}\widehat{r}^P_{nw}(x)+\sqrt{n^Qh^Q}\widehat{r}^Q_{nw}(x)}{\sqrt{n^Ph^P}+\sqrt{n^Qh^Q}},
  $$
  where $\widehat{r}^P_{nw}(x)$ and $\widehat{r}^Q_{nw}(x)$ are the N-W estimators for P-model and Q-model, respectively.
  \item[3)] The data-driven weighted average of the N-W estimator denoted by WA and defined by
  $$
  \widehat{r}^Q_{WA}(x)=w^P\widehat{r}^P_{nw}(x)+w^Q\widehat{r}^Q_{nw}(x),
  $$
  where $w^P$ and $w^Q$ are chosen by the criterion
  $$
    (w^P, w^Q)^\top={\arg\min}_{w^P+w^Q=1 \atop w^P, w^Q \in [0,1]}\sum_{i=1}^{n^Q}(Y^Q_i-\widehat{r}^Q_{WA(-i)}(X_i^Q))^2
  $$
  with $\widehat{r}^Q_{WA(-i)}(X_i^Q)$ being the leave-one-out version  of the WA.
\end{itemize}
The estimation performance is measured with the mean integrated squared error (MISE) derived by 1000 replications. All the kernel estimators are constructed by the Gaussian kernel and the bandwidth is chosen by CV criterion defined in (2.9).

\subsection{Identical source problem}
\begin{figure*}[t]
\centering
\includegraphics[width=0.8\textwidth]{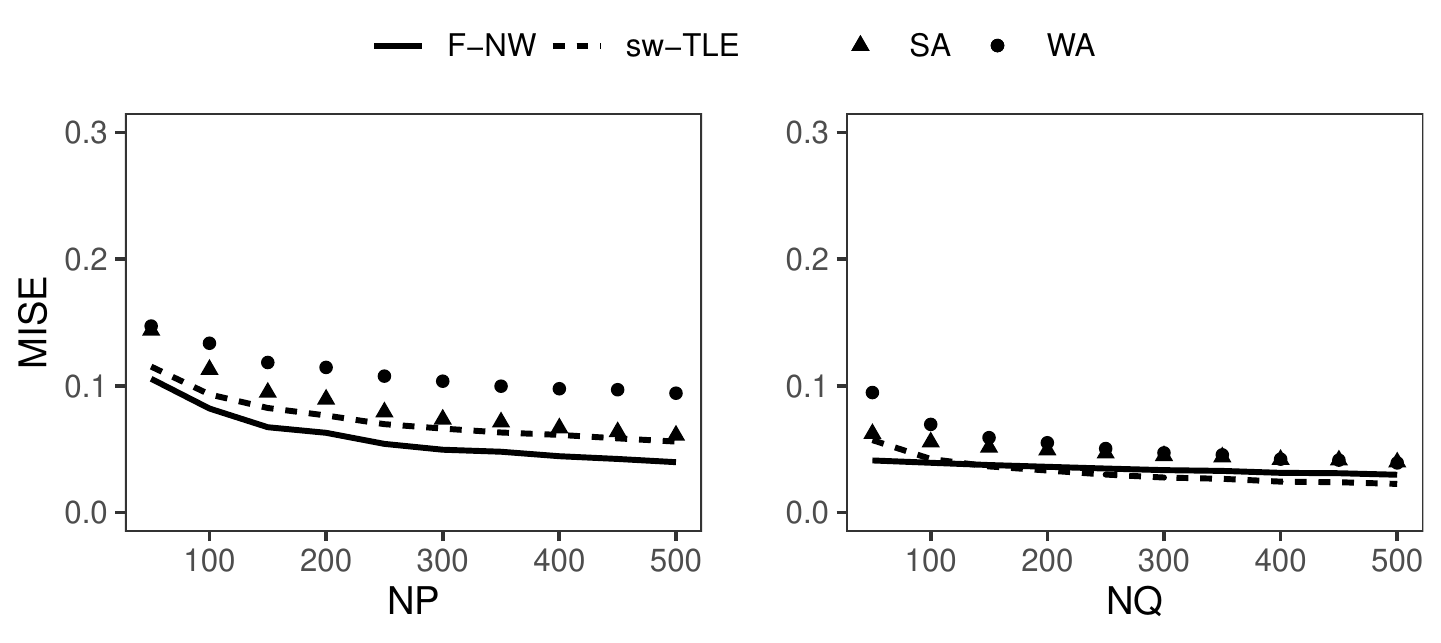}
\caption{\small The MISE curves of the estimators in the identical source problem. In the left subfigure, $n^Q=50$ is fixed with  varying $n^P$. In the right subfigure, $n^P=500$ is fixed with varying $n^Q$}
\label{fig:e1}
\end{figure*}

We first consider the case where the source model P and the target model Q are identical:
\begin{align*}
  \text{Model }P: ~~&Y^P=\cosh(X^P)+\varepsilon^P,\\
  \text{Model }Q: ~~&Y^Q=\cosh(X^Q)+\varepsilon^Q,
\end{align*}
where independent random variables $X^P\sim U[-2,2]$, $X^Q\sim U[-2,2]$, $\varepsilon^P\sim N(0,0.2^2)$ and $\varepsilon^Q\sim N(0,0.2^2)$. In this example, the N-W regression utilizes all the data from P-model and Q-model, denoted as F-NW, as shown above. The MISE curves are given in Fig. \ref{fig:e1} and the detailed simulation data can be found in Table \ref{tab:e1}. We have the following findings:
\begin{enumerate}
  \item The MISE decreases with increased $n^P$ or $n^Q$ for all the methods.
  \item Our sw-TLE is much better than the SA and the WA  in the sense that MISE of sw-TLE estimation is significantly smaller than those of the SA and  the WA.
  \item Like the argument in Remark 2.4., our method is even better than the full data N-W regression when P = Q and $n^P/n^Q$ are not too large.
\end{enumerate}
\begin{table}
\caption{\label{tab:e1} \small The MISE of the sw-TLL, the F-NW, the SA and the WA in the identical source case with fixed $n^Q=50$ varying $n^P$ and fixed $n^P=500$ varying $n^Q$.}
\tiny
\centering
\begin{tabular}{ccccccccccc }
\hline
\multicolumn{5}{c}{$n^P=500$, varying $n^Q$}&&\multicolumn{5}{c}{$n^Q=50$, varying $n^P$}\\
\cline{1-5}
\cline{7-11}
$n^Q$&F-NW&sw-TLE&SA&WA&&$n^P$&F-NW&sw-TLE&SA&WA\\
\cline{1-5}
\cline{7-11}
10&0.042 &0.176&0.095 &0.536&&50&0.117 &0.117 &0.146 &0.150\\
50&0.040 &0.058 &0.062 &0.096&&100&0.083 &0.096 &0.115 &0.136\\
100&0.038 &0.042 &0.055 &0.069&&200&0.061 &0.077 &0.089 &0.117\\
200&0.035 &0.032 &0.048 &0.054&&500&0.040 &0.058 &0.062 &0.096\\
500&0.029 &0.022 &0.039 &0.039&&1000&0.029 &0.047 &0.046 &0.086\\
\hline
\end{tabular}
\end{table}

\subsection{Unrelated source problem}
\begin{figure*}[t]
\centering
\includegraphics[width=0.8\textwidth]{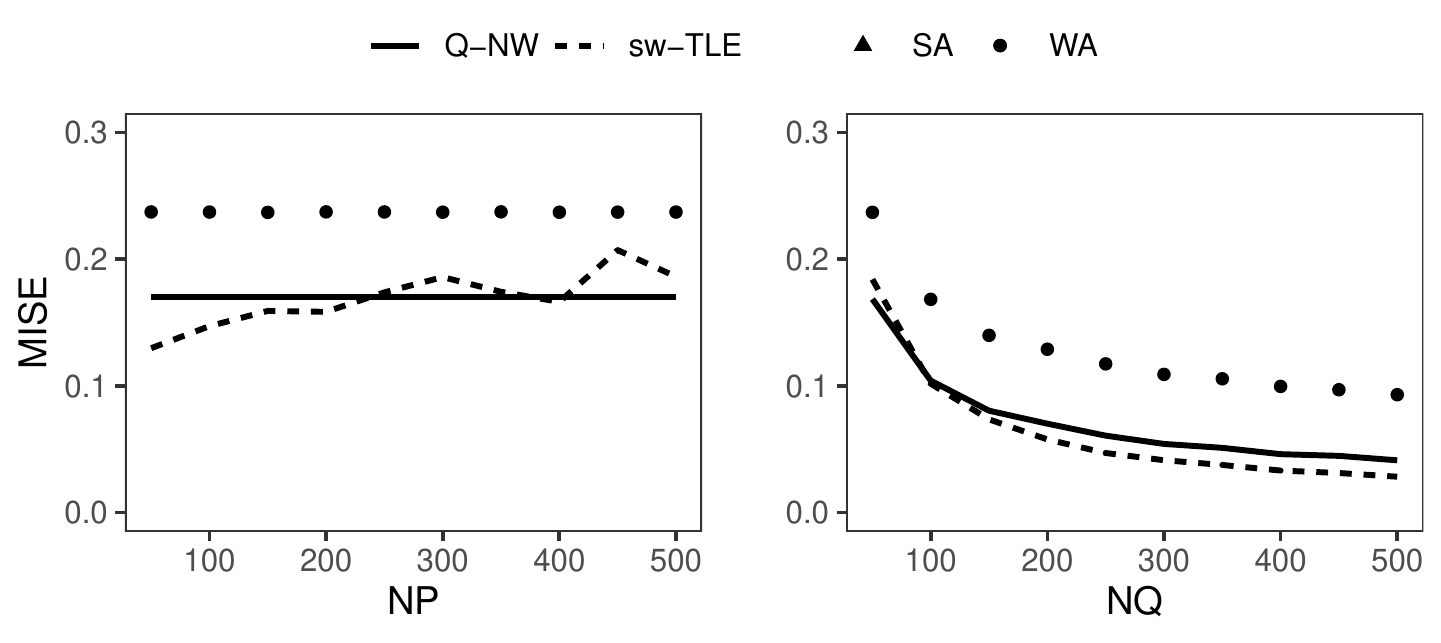}
\caption{\small The MISE curves of the estimators in the unrelated source problem. In the left subfigure, $n^Q=50$ is fixed with varying $n^P$; in the right subfigure, $n^P=500$ is fixed with varying $n^Q$. Note that the curves of MISE of Q-NW are beyond the realm of the figure.}
\label{fig:e3}
\end{figure*}

Our method is of adaptability to the unrelated source models. Here we use an example to illustrate this characteristic. Consider the following completely different models:
\begin{align*}
  \text{Model }P: ~~&Y^P=0.5+\varepsilon^P,\\
  \text{Model }Q: ~~&Y^Q=\cosh(X^Q)+\varepsilon^Q,
\end{align*}
where independent random variables $X^P\sim U[-2,2]$, $X^Q\sim U[-2,2]$, $\varepsilon^P\sim N(0,0.2^2)$ and $\varepsilon^Q\sim N(0,0.2^2)$.
In this example, the sw-TLE of $r^Q(x)$ is constructed by (2.13).
The MISE curves
are given in Fig. \ref{fig:e3} and the detailed numerical results can be found in Table \ref{tab:e3}. We have the following findings:
\begin{enumerate}
  \item The sw-TLE method works better than the WA because its MISE is smaller. The MISE of the SA is much larger than 1, see Table \ref{tab:e3}, implying that it dose not suit for treating this type of problem.
  \item When $n^P/n^Q$ is not too large, the sw-TLE is better than the NW. The performance of the sw-TLE becomes worse when  $n^P/n^Q$ is too large.
\end{enumerate}
\begin{table}
\caption{\label{tab:e3} \small The MISE of the sw-TLL, the Q-NW, the SA and the WA in the identical source cases with fixed $n^Q=50$ varying $n^P$ and fixed $n^P=500$ varying $n^Q$.}
\tiny
\centering
\begin{tabular}{ccccccccccc }
\hline
\multicolumn{5}{c}{$n^P=500$, varying $n^Q$}&&\multicolumn{5}{c}{$n^Q=50$, varying $n^P$}\\
\cline{1-5}
\cline{7-11}
$n^Q$&Q-NW&sw-TLE&SA&WA&&$n^P$&NW&sw-TLE&SA&WA\\
\cline{1-5}
\cline{7-11} \tiny
10&1.105 &1.318 &7.647 &1.119&&50&0.177 &0.091 &2.988 &0.242\\
50&0.177 &0.123 &5.642 &0.246 &&100&0.177 &0.109 &3.745 &0.247\\
100&0.106 &0.053 &4.753 &0.170 &&200&0.177 &0.113 &4.561 &0.247\\
200&0.070 &0.027 &3.872 &0.128 &&500&0.177 &0.123 &5.642 &0.246\\
500&0.042 &0.012 &2.790 &0.093 &&1000&0.177 &0.145 &6.429 &0.245\\
\hline
\end{tabular}
\end{table}

\subsection{Multi-source problem}
Our method works well in the multi-source case. Set
\begin{align*}
  \text{Model }P_1: ~~&Y^{P}_1=0.5\exp\left(\frac{(X^{P}_1)^2}{2}\right)+\varepsilon^{P}_1,\\
  \text{Model }P_2: ~~&Y^{P}_2=1-\cos\left(X^{P}_2\right)+\varepsilon^{P}_2,\\
  \text{Model }Q: ~~&Y^Q=\cosh(X^Q)+\varepsilon^Q,
\end{align*}
where independent random variables $X^{P}_1\sim U[-2,2]$, $X^{P}_2\sim U[-2,2]$, $X^Q\sim U[-2,2]$, $\varepsilon^{P}_1\sim N(0,0.2^2)$, $\varepsilon^{P}_2\sim N(0,0.2^2)$ and $\varepsilon^Q\sim N(0,0.2^2)$. In multi-source problem, our method needs to choose the weights as in \eqref{equ_muti} by the criterion
$$
    (w_1, w_2)^\top={\arg\min}_{w_1+w_2=1 \atop w_1, w_2 \in [0,1]}\sum_{i=1}^{n^Q}(Y^Q_i-\widehat{r}^Q_b(X_i^Q; w_1,w_2))^2,
$$
where
$\widehat{r}^Q_b(x; w_1,w_2)=w_1\widehat{r}^{P}_1(x)\widehat{\eta}_1(x)+w_2\widehat{r}^{P}_2(x)\widehat{\eta}_2(x)$.
The MISE curves with the same source data size are presented in Fig. \ref{fig:e4}. Other data settings with different size and the detailed simulation data can be found in Table \ref{tab:e4}. We have the following observations:
\begin{enumerate}
  \item The sw-TLE method works better than the WA and the NW due to the smaller value of the MISE. The MISE of SA is significantly larger than 1, see Table \ref{tab:e4}, indicating that the SA dose not suit for treating the unrelated multi-source problem.
  \item As the increasing of $n^Q$, the performance of NW becomes better, closing to the sw-TLE.
  \item The difference between sizes of two source data is influential in the performance of the sw-TLE, see Table \ref{tab:e4}.
\end{enumerate}
\begin{figure*}[t]
\centering
\includegraphics[width=0.8\textwidth]{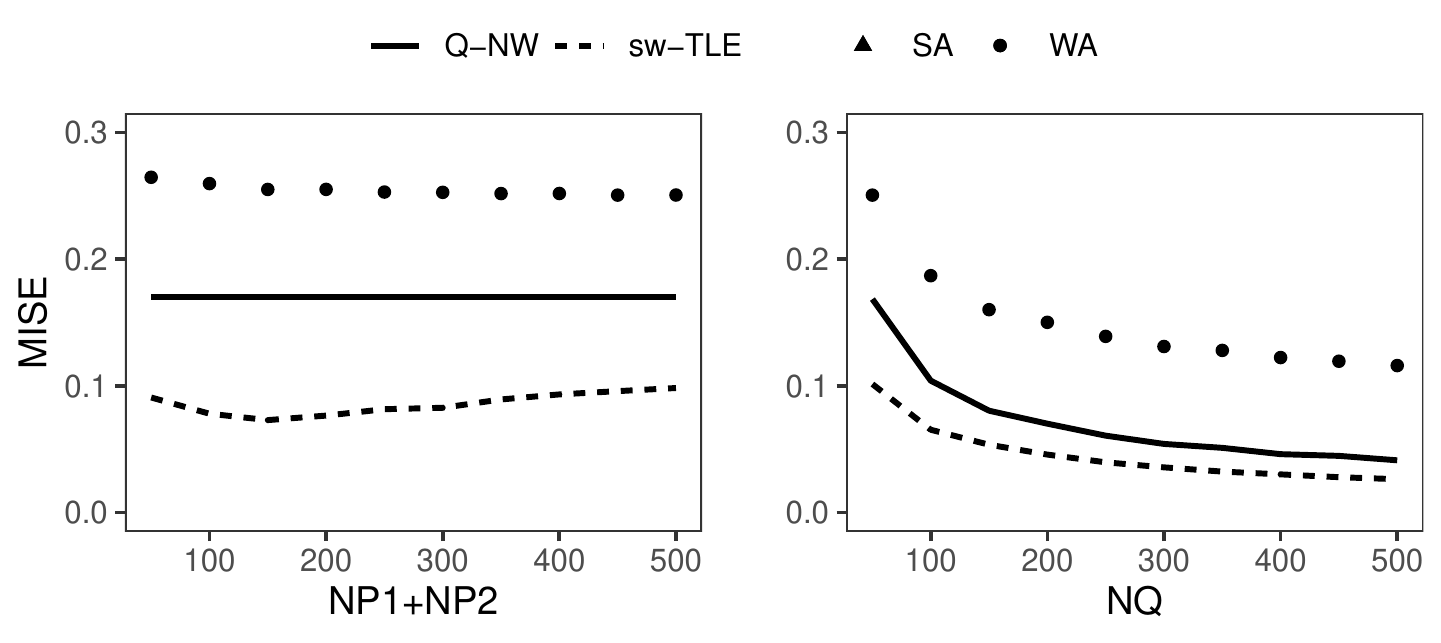}
\caption{\small The MISE curves of the estimators in the muti-source problem. In the right subfigure, $n^P=500$ is fixed with varying $n^Q$. Note that the curves of the MISE of Q-NW are beyond the realm of the figure.}
\label{fig:e4}
\end{figure*}

\begin{table}[thp]
\caption{\label{tab:e4}\small The MISE of the sw-TLL, the Q-NW, the SA and the WA in the similar source case with fixed $n^{P_1}+n^{P_2}=500$ varying $n^Q$ and fixed $n^Q=50$ varying $n^{P_1}$, $n^{P_2}$.}
\tiny
\centering
\begin{tabular}{ccccccccccc }
\hline
\multicolumn{5}{c}{$n^{P_1}=100$, {$n^{P_2}=400$} varying $n^Q$}&&\multicolumn{5}{c}{$n^Q=50$, varying $n^{P_1}$, $n^{P_2}$}\\
\cline{1-5}
\cline{7-11}
$n^Q$&Q-NW&sw-TLE&SA&WA&&$n^{P_1}$, $n^{P_2}$&NW&sw-TLE&SA&WA\\

\cline{1-5}
\cline{7-11}
10&1.050 &0.412 &4.041 &0.954 &&10, 40&0.177&	0.100 &2.239 &0.247\\
50&0.177 &0.121 &3.214 &0.219 &&20, 80&0.177&	0.098 &2.477 &0.234\\
100&0.106 &0.086 &2.824 &0.161 &&40, 160&0.177 &0.095 &2.801 &0.226\\
200&0.070 &0.059 &2.400 &0.128 &&100, 400&0.177 &0.033 &3.214 &0.219\\
500&0.042 &0.035 &1.848 &0.100 &&200, 800&0.177 &0.014 &3.504 &0.218\\
\hline
\hline
\multicolumn{5}{c}{$n^{P_1}=250$, {$n^{P_2}=250$} varying $n^Q$}&&\multicolumn{5}{c}{$n^Q=50$, varying $n^{P_1}$, $n^{P_2}$}\\
\cline{1-5}
\cline{7-11}
$n^Q$&Q-NW&sw-TLE&SA&WA&&$n^{P_1}$, $n^{P_2}$&NW&sw-TLE&SA&WA\\
\cline{1-5}
\cline{7-11}
10&1.050 &0.370 &3.375 &1.013 &&25, 25&0.177 &0.080 &1.856 &0.233\\
50&0.177 &0.099 &2.705 &0.234 &&50, 50&0.177 &0.065 &2.105 &0.228\\
100&0.106 &0.066 &2.386 &0.168& &100, 100&0.177 &0.062 &2.365&0.221\\
200&0.070 &0.044 &2.043 &0.131 &&250, 250&0.177 &0.020 &2.704 &0.216\\
500&0.042 &0.024 &1.582 &0.098 &&500, 500&0.177 &0.010 &2.934 &0.216\\
\hline
\hline
\multicolumn{5}{c}{$n^{P_1}=400$, {$n^{P_2}=100$} varying $n^Q$}&&\multicolumn{5}{c}{$n^Q=50$, varying $n^{P_1}$, $n^{P_2}$}\\
\cline{1-5}
\cline{7-11}
$n^Q$&Q-NW&sw-TLE&SA&WA&&$n^{P_1}$, $n^{P_2}$&NW&sw-TLE&SA&WA\\
\cline{1-5}
\cline{7-11}
10&1.050 &0.158 &2.723 &0.947 &&40, 10&0.177 &0.081 &1.516 &0.235\\
50&0.177 &0.043 &2.157 &0.215 &&80, 20&0.177 &0.052 &1.685 &0.228\\
100&0.106 &0.031 &1.897 &0.158 &&160, 40&0.177 &0.040 &2.412&	0.619\\
200&0.070 &0.022 &1.615 &0.125&&	400, 100&0.177 &0.013 &2.157 &0.215\\
500&0.042 &0.014 &1.242 &0.098&&	800, 200&0.177 &0.007 &2.346 &0.215\\
\hline
\end{tabular}
\end{table}

\section{Some details of numerical studies in main text}
Here, we give the MISE tables of estimations in the similar source problem mentioned in Section 4.1 and the data figure for Section 4.3.
\begin{table}[thp]
\caption{\label{tab:e2_1}\small The MISE of the sw-TLL, the Q-NW, the SA and the WA in the similar source case with fixed $n^P=500$  and varying $n^Q$.}
\tiny
\centering
\begin{tabular}{ccccccccccc }
\hline
\multicolumn{11}{c}{$b=1$}\\
\hline
\multirow{2}{*}{$n^Q$}&\multirow{2}{*}{Q-NW}&\multicolumn{3}{c}{$a=1$}&\multicolumn{3}{c}{$a=2$}&\multicolumn{3}{c}{$a=3$}\\
&&sw-TLE&SA&WA&sw-TLE&SA&WA&sw-TLE&SA&WA\\
\hline
10&1.050 &0.120 &1.029 &0.766 &0.173 &6.633 &0.889&	0.485 &17.845 &0.898\\
50&0.177 &0.072 &0.794 &0.138 &0.074 &5.014 &0.156 &0.077 &13.429 &0.185\\
100&0.106 &0.041 &0.665 &0.084 &0.041 &4.205 &0.097 &0.042 &11.268&	0.128\\
200&0.070 &0.023 &0.540 &0.057&	0.024 &3.409 &0.069 &0.024&	9.139 &0.102\\
500&0.042 &0.011 &0.388 &0.035 &0.011 &2.442 &0.049 &0.012 &6.546 &0.084\\

\hline
\hline
\multicolumn{11}{c}{$b=2$}\\
\hline
\multirow{2}{*}{$n^Q$}&\multirow{2}{*}{Q-NW}&\multicolumn{3}{c}{$a=1$}&\multicolumn{3}{c}{$a=2$}&\multicolumn{3}{c}{$a=3$}\\
&&sw-TLE&SA&WA&sw-TLE&SA&WA&sw-TLE&SA&WA\\
\hline
10&1.050 &0.175 &16.067 &0.785 &0.193 &29.415 &0.804 &0.109 &48.372 &0.839\\
50&0.177 &0.070 &12.131 &0.134 &0.073 &22.145 &0.171 &0.071 &36.353 &0.229\\
100&0.106 &0.040 &10.175 &0.088 &0.041 &18.581 &0.127 &0.040 &30.511 &0.186\\
200&0.070 &0.023 &8.247 &0.069&	0.023 &15.065 &0.110 &0.023 &24.745 &0.171\\
500&0.042&0.011 &5.901 &0.058 &0.011 &10.786 &0.101 &0.011 &17.721 &0.164\\

\hline
\hline
\multicolumn{11}{c}{$b=3$}\\
\hline
\multirow{2}{*}{$n^Q$}&\multirow{2}{*}{Q-NW}&\multicolumn{3}{c}{$a=1$}&\multicolumn{3}{c}{$a=2$}&\multicolumn{3}{c}{$a=3$}\\
&&sw-TLE&SA&WA&sw-TLE&SA&WA&sw-TLE&SA&WA\\
\hline
10&1.050 &0.331 &49.585 &0.743 &0.387 &70.678 &0.785 &0.390 &97.380 &0.848\\
50&0.177 &0.064 &37.296 &0.189 &0.061 &53.102 &0.254 &0.065 &73.104 &0.340\\
100&0.106 &0.036 &31.297 &0.157 &0.039 &44.569 &0.224 &0.037 &61.366 &0.311\\
200&0.070 &0.023 &25.379 &0.149 &0.023 &36.148 &0.218 &0.023 &49.777 &0.307\\
500&0.042 &0.011 &18.173 &0.149 &0.011 &25.889 &0.220 &0.011 &35.655 &0.312\\

\hline
\end{tabular}
\end{table}

\begin{table}[thp]
\caption{\label{tab:e2_2}\small The MISE of the sw-TLL, the Q-NW, the SA and the WA in the similar source case with fixed $n^Q=50$  and varying $n^P$.}
\tiny
\centering
\begin{tabular}{ccccccccccc }
\hline
\multicolumn{11}{c}{$b=1$}\\
\hline
\multirow{2}{*}{$n^P$}&\multirow{2}{*}{Q-NW}&\multicolumn{3}{c}{$a=1$}&\multicolumn{3}{c}{$a=2$}&\multicolumn{3}{c}{$a=3$}\\
&&sw-TLE&SA&WA&sw-TLE&SA&WA&sw-TLE&SA&WA\\
\hline
50&0.177 &0.082 &0.388 &0.035 &0.084 &2.442 &0.049 &0.085 &6.546 &0.084\\
100&0.177 &0.078 &0.470 &0.144 &0.079 &3.077 &0.161 &0.082 &8.339 &0.190\\
200&0.177 &0.074 &0.594 &0.140 &0.076 &3.879 &0.154 &0.079 &10.473 &0.182\\
500&0.177 &0.072 &0.794 &0.138 &0.074 &5.014 &0.156 &0.077 &13.429 &0.185\\
1000&0.177 &0.070 &0.947 &0.135 &0.072 &5.845 &0.154 &0.076 &15.582 &0.184\\
\hline
\hline
\multicolumn{11}{c}{$b=2$}\\
\hline
\multirow{2}{*}{$n^P$}&\multirow{2}{*}{Q-NW}&\multicolumn{3}{c}{$a=1$}&\multicolumn{3}{c}{$a=2$}&\multicolumn{3}{c}{$a=3$}\\
&&sw-TLE&SA&WA&sw-TLE&SA&WA&sw-TLE&SA&WA\\
\hline
50&0.177 &0.084&5.213 &0.139 &0.084 &9.983 &0.175&	0.086 &16.804 &0.232\\
100&0.177&	0.078 &7.085 &0.139 &0.076 &13.326&	0.176 &0.079 &22.222 &0.233\\
200&0.177 &0.075 &9.174 &0.132 &0.074 &17.009 &0.167 &0.075 &28.152 &0.223\\
500&0.177 &0.072 &12.131 &0.134 &0.072&	22.145 &0.171 &0.073 &36.353 &0.229\\
1000&0.177 &0.070 &14.326 &0.132 &0.071 &25.921 &0.169 &0.071 &42.356 &0.227\\
\hline
\hline
\multicolumn{11}{c}{$b=3$}\\
\hline
\multirow{2}{*}{$n^P$}&\multirow{2}{*}{Q-NW}&\multicolumn{3}{c}{$a=1$}&\multicolumn{3}{c}{$a=2$}&\multicolumn{3}{c}{$a=3$}\\
&&sw-TLE&SA&WA&sw-TLE&SA&WA&sw-TLE&SA&WA\\
\hline
50&0.177 &0.086 &16.351 &0.191 &0.085 &23.912 &0.255 &0.087 &33.523 &0.340\\
100&0.177 &0.084 &22.082 &0.191 &0.083 &31.958 &0.256 &0.085 &44.487 &0.341\\
200&0.177 &0.081 &28.427 &0.182 &0.081 &40.812 &0.246 &0.080 &56.506 &0.330\\
500&0.177 &0.080 &37.296 &0.189 &0.080 &53.102 &0.254 &0.078 &73.104 &0.340\\
1000&0.177&	0.079 &43.850 &0.187 &0.079 &62.143 &0.252 &0.078 &85.275 &0.338\\
\hline
\end{tabular}
\end{table}
\begin{figure*}[t]
\centering
\includegraphics[width=0.8\textwidth]{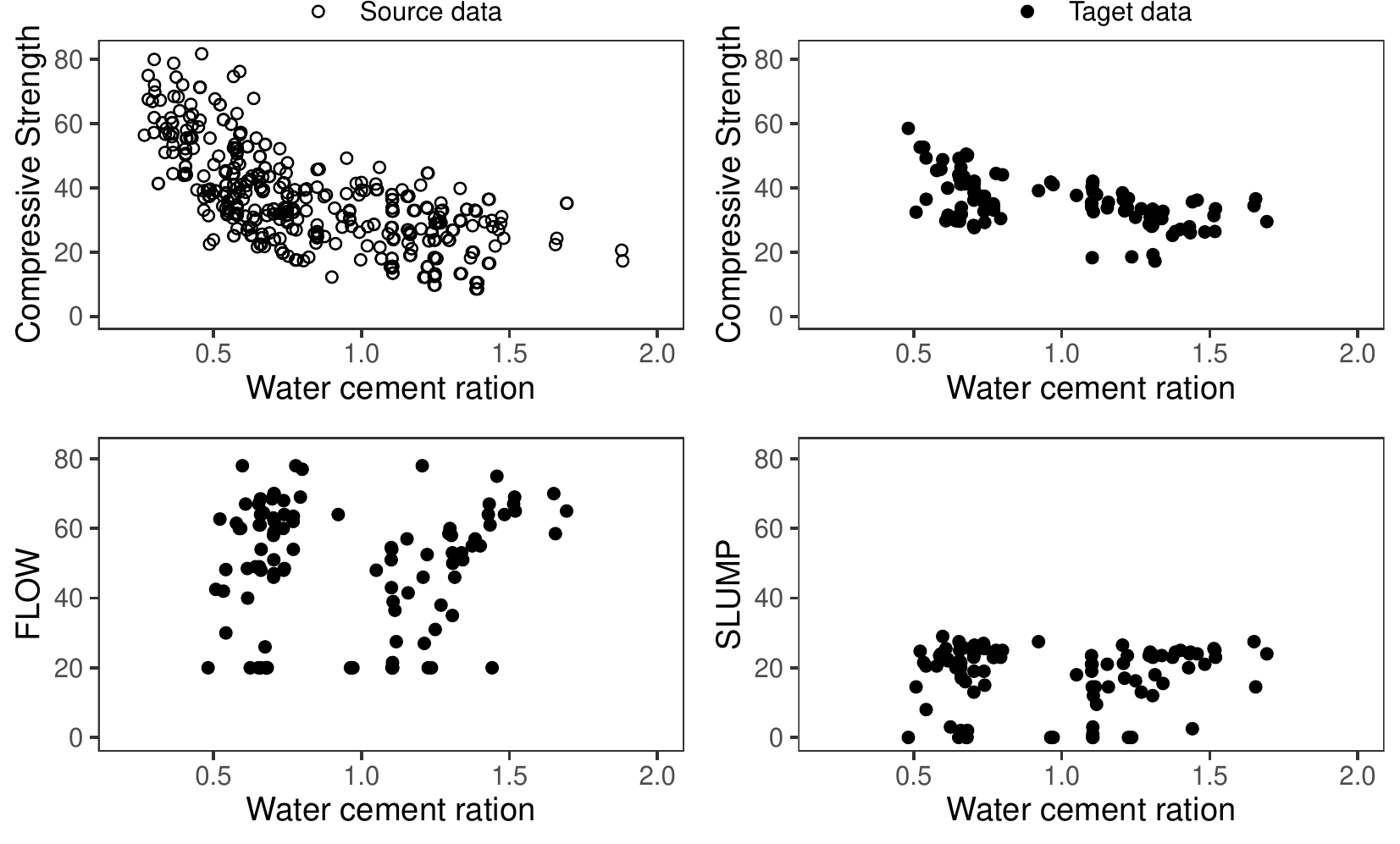}
\caption{\small The real data sets used in Section 4.3.}
\label{fig:e4}
\end{figure*}

\end{document}